\documentclass[11pt,a4paper]{article}

\usepackage{jheppub}
\usepackage{multirow}
\usepackage{ulem}

\title{$Z_N$ twisted orbifold models with magnetic flux}

\author[a]{Tomo-hiro~Abe,}
\author[b]{Yukihiro~Fujimoto,}
\author[a]{Tatsuo~Kobayashi,}
\author[b]{Takashi~Miura,}
\author[c]{Kenji~Nishiwaki}
\author[b]{and Makoto~Sakamoto}

\affiliation[a]{Department of Physics, Kyoto University, Kyoto 606-8502, Japan}
\affiliation[b]{Department of Physics, Kobe University, Kobe 657-8501, Japan}
\affiliation[c]{Regional Centre for Accelerator-based Particle Physics, \\
Harish-Chandra Research Institute, Allahabad 211 019, India}

\emailAdd{t-abe@scphys.kyoto-u.ac.jp}
\emailAdd{093s121s@stu.kobe-u.ac.jp}
\emailAdd{kobayash@gauge.scphys.kyoto-u.ac.jp}
\emailAdd{takashi.miura@people.kobe-u.ac.jp}
\emailAdd{nishiwaki@hri.res.in}
\emailAdd{dragon@kobe-u.ac.jp}

\abstract{
We propose new backgrounds of extra dimensions to lead to 
four-dimensional chiral models with three generations of matter
fermions, that is $T^2/Z_N$ twisted orbifolds with magnetic fluxes.
We consider gauge theory on six-dimensional space-time, which contains 
the $T^2/Z_N$  orbifold with magnetic flux, Scherk-Schwarz phases 
and Wilson line phases.
We classify all the possible Scherk-Schwarz and Wilson line phases on 
$T^2/Z_N$  orbifolds with magnetic fluxes.
The behavior of zero modes is studied.
We derive the number of zero modes for each eigenvalue of 
the $Z_N$ twist, showing explicitly examples of wave functions.
We also investigate  Kaluza-Klein mode functions and mass 
spectra.
}


\arxivnumber{1309.4925}

\begin{document}

\maketitle

\section{Introduction}

Extra-dimensional field theories play an important role in 
particle physics.
Phenomenological aspects of four-dimensional low energy 
effective field theory strongly depend on geometrical aspects 
of compactification of extra dimensions and some gauge backgrounds 
as well as other backgrounds.
For example, one of the simplest compactifications is 
a torus compactification.
However, the toroidal compactification without any 
non-trivial gauge backgrounds leads to a four-dimensional 
non-chiral theory, and that is not realistic.
In general a more complicated geometrical background can 
lead to a four-dimensional chiral theory, but 
it is difficult to solve zero-mode equations 
in generic background and derive the four-dimensional 
low energy effective field theory.

Torus compactifications with some magnetic fluxes 
are quite interesting extra-dimensional
 backgrounds \cite{Bachas:1995ik,Cremades:2004wa}.\footnote{
See for string magnetized D-brane models \cite{Blumenhagen:2005mu} 
and references therein.}
That can realize chiral spectra in four-dimensional 
low energy effective field theory.
One can solve zero-mode equations analytically 
and their zero-mode profiles are nontrivially quasi-localized.

The number of zero modes depends on 
the magnitude of magnetic flux, 
and three-generation chiral models can be obtained 
by choosing properly magnetic fluxes.\footnote{
Five-dimensional models with point interactions are 
another type of interesting approach to realize 
three generations \cite{Fujimoto:2012}.
}
In addition, since zero modes are quasi-localized, 
their couplings in the four-dimensional low energy effective 
field theory are 
non-trivial.
That is, when they are quasi-localized far away from 
each other, their couplings can be suppressed.
Thus, magnetic flux backgrounds are quite interesting.
Indeed, several studies have been carried out 
to derive four-dimensional realistic models 
and study their phenomenological aspects, 
e.g., Yukawa couplings \cite{Cremades:2004wa},\footnote{
Within the framework of superstring theory, 
magnetized D-brane models are T-dual of intersecting D-brane 
models \cite{Blumenhagen:2005mu}.
Yukawa couplings are also computed in intersecting D-brane models 
\cite{Cvetic:2003ch}.} 
realization of quark/lepton masses 
and their mixing angles \cite{Abe:2012fj}, 
higher order couplings \cite{Abe:2009dr}, 
flavor symmetries \cite{Abe:2009vi,Abe:2009uz,BerasaluceGonzalez:2012vb,Marchesano:2013ega,Honecker:2013hda}, 
massive modes \cite{Hamada:2012wj}, and so on 
\cite{Sakamoto:2003,Antoniadis:2004pp,Choi:2009pv,DiVecchia:2011mf,Abe:2012ya,DeAngelis:2012jc,Abe:2013bba}.

The $T^2/Z_2$ twisted orbifold compactification with 
magnetic flux is also interesting \cite{Abe:2008fi,Abe:2008sx}.\footnote{
See for heterotic models on magnetized orbifolds
\cite{Nibbelink:2012de} and also for shifted $T^2/Z_N$ orbifold models
with magnetic flux \cite{Fujimoto:2013xha}.}
The zero modes and their wave functions on $T^2$ are 
classified by $Z_2$ charges, that is, $Z_2$ even and odd.
Then, either $Z_2$ even or odd eigenstates are 
projected out exclusively by the orbifold boundary conditions.
Thus, the number of zero modes on  $T^2/Z_2$ with 
magnetic flux is different from one on $T^2$ with the same magnetic flux.
Also each of zero-mode wave functions on $T^2/Z_2$ can be 
derived analytically and it has a non-trivial profile.
One can construct three-generation orbifold models, 
which are different from models on $T^2$ with magnetic flux.
Such analysis can be extended into higher dimensional models 
such as $T^6/Z_2$ and $T^6/(Z_2 \times Z'_2)$ orbifolds 
with magnetic fluxes \cite{Abe:2008fi,Abe:2008sx}.

In addition to $T^2/Z_2$, there are other two-dimensional orbifolds, 
$T^2/Z_N$ for $N=3,4,6$ \cite{Dixon:1985jw}.
(See for their geometrical aspects 
\cite{Katsuki:1989bf,Kobayashi:1991rp,Choi:2006qh}.\footnote{
In a higher-dimensional field theory without magnetic flux, detailed studies of the $SU(N)$ and $SO(N)$ gauge theory have been made in ref.\cite{Kawamura:2008mz,Kawamura:2009sa,Kawamura:2007,Kawamura:2009,Goto:2013jma}.
Especially, some candidates of three-generation models have been shown in ref.\cite{Goto:2013jma}.})
Moreover, there are various orbifolds in six dimensions 
like $T^6/Z_7, T^6/Z_8, T^6/Z_{12}$, etc.
Obviously, geometrical aspects of $T^2/Z_N$ with $N=3,4,6$ 
are different from those of $T^2/Z_2$.
Thus, one can derive interesting models on $T^2/Z_N$ for $N=3,4,6$ 
with magnetic fluxes, which are different from those 
on $T^2/Z_2$.
Hence, it is our purpose to study these orbifold models 
with magnetic fluxes.
In addition, non-trivial (discrete) Scherk-Schwarz phases
\cite{Scherk:1978ta} and 
Wilson line phases are possible on orbifolds 
\cite{Ibanez:1986tp,Kobayashi:1990mi,Kobayashi:1991rp}.\footnote{
Also in intersecting D-brane models, Scherk-Schwarz phases were 
discussed in \cite{Angelantonj:2005hs} and discrete Wilson lines 
were studied in \cite{Blumenhagen:2005tn} (see also \cite{Angelantonj:2009yj}).}
Such backgrounds have not been taken into account 
in \cite{Abe:2008fi,Abe:2008sx} for the study on 
$T^2/Z_2$ with magnetic flux.
Here, we also consider these phases.

In this paper, we study $T^2/Z_N$ orbifold models for $N=2,3,4,6$ 
with magnetic flux, Scherk-Schwarz phases and Wilson lines.
We clarify possible Scherk-Schwarz phases as well as 
Wilson lines on $T^2/Z_N$ orbifolds for $N=2,3,4,6$ 
with magnetic fluxes.
Then, we study the behavior of zero modes on $T^2/Z_N$ 
for each eigenvalue under the $Z_N$ twist.
We will show that one can obtain three-generation models 
in various cases and model building becomes rich.
Furthermore, we show Kaluza-Klein mode functions and its interesting mass spectrum.

This paper is organized as follows.
In section~\ref{sec:RevU1T2}, we review the $U(1)$ gauge theory on a two-dimensional torus 
with magnetic fluxes.
In section~\ref{sec:Twi_orbi w/mf}, we study the general formalism of $Z_N$ twisted orbifolds
with magnetic flux.
Especially, we investigate the form of the eigenfunctions 
for each  $Z_N$ eigenvalue and the allowed values of 
important parameters such as Scherk-Schwarz phases and Wilson lines 
on each orbifold with magnetic flux.
In section~\ref{sec:0mode eigenstates}, we analyze the number of zero-mode eigenfunctions, 
that is, the number of generations for matter fermions on each orbifold.
In section~\ref{sec:KK and masses}, we also show Kaluza-Klein mode functions and their mass spectrum.
Section~\ref{sec:Conclusion} is devoted to the conclusions and discussions.
In appendix \ref{LsandGm} our notation is summarized.
In appendix \ref{Wlp&alp} we show the relations between Wilson lines 
and Scherk-Schwarz phases.
In appendix \ref{Exaple of calculation} we show some examples 
of calculations on zero-mode wave functions on 
$T^2/Z_N$ with magnetic fluxes.

\section{Gauge field theory on $M^4\times T^2$ with magnetic flux \label{sec:RevU1T2}}

Let us study the behavior of gauge and matter fields on
six-dimensional space-time, which contains four-dimensional Minkowski
space-time $M^4$ and an extra two-dimensional torus $T^2$.
We denote coordinates on $M^4$ by $x^\mu$  ($\mu=0,1,2,3$) and 
we use the complex coordinate $z$ on $T^2$. 
We consider a theory containing the torus with magnetic flux.
Then, one can obtain an attractive feature that 
chiral zero-mode fermions appear and their 
number is determined by the magnitude of the magnetic flux.
We will see it below.

First of all, we consider the Lagrangian density based on a $U(1)$ gauge theory on $M^4\times T^2$ such as
\begin{align}
\mathcal{L}_{\mathrm{6D}}=
&-{1\over 4}F^{MN}F_{MN}+i\bar{\Psi}_+\Gamma^MD_M \Psi_+ ,
\label{6DLagrangian}
\end{align}
where $M,N=\mu (=0,1,2,3), z,\bar{z}$ and $D_M=\partial_M-iqA_M(x,z)$~\footnote{
In this paper, we use a notation as in appendix \ref{LsandGm}.
Note that fields such as $A_M(x,z)$, $A_z^{(b)}(z)$,
$\Psi_{\pm}(x,z)$, $\psi_{\pm,n}(z)$ and so on are written by
functions depending on not only $z$ 
but also $\bar{z}$.}
with a $U(1)$ charge $q$.
Here, $\Psi_{\pm}$ are six-dimensional Weyl fermions, and are obtained by projection operators ${1\pm \Gamma_7 \over 2}$ such as
\begin{align}
&\hspace{-15.4mm}\Psi_{\pm} \equiv {1\pm \Gamma_7 \over 2}\Psi,~~\notag \\
\Psi_{+}(x,z)&=\psi_{R}(x,z) +\psi_{L}(x,z) \notag \\
&=\sum_n\left(\psi_{4R,n}(x) \otimes \psi_{2+,n}(z) + \psi_{4L,n}(x) \otimes \psi_{2-,n}(z) \right),~~\notag \\
\Psi_{-}(x,z)&=\psi'_{L}(x,z) + \psi'_{R}(x,z) \notag \\
&=\sum_n\left(\psi'_{4L,n}(x) \otimes \psi_{2+,n}(z) + \psi'_{4R,n}(x) \otimes \psi_{2-,n}(z)\right),
\label{6D_Dirac to two Weyl}
\end{align}
where $n$ means the label of mass eigenstates.
$\Psi$ is a six-dimensional Dirac fermion, $\psi_{4R/L,n}$ and $\psi'_{4R/L,n}$ are 
four-dimensional right/left-handed fermions, 
and $\psi_{2\pm,n}$ are two-dimensional Weyl fermions.
For convenience, we also use the following notation:
\begin{align}
\psi_{2+,n}(z)=\left(
\begin{array}{c}
\psi_{+,n} (z)\\ 0
\end{array}
\right), ~~~
\psi_{2-,n}(z)=\left(
\begin{array}{c}
0\\ \psi_{-,n}(z)
\end{array}
\right).
\end{align}
Then, the action for the one Weyl fermion $\Psi_+$ and its gauge interaction can be written as
\begin{align}
S^{\mathrm{Weyl}}
&=\int_{M^4} d^4x\int_{T^2} dzd\bar{z} ~i\bar{\Psi}_+\Gamma^MD_M
\Psi_+ \notag \\
&=\int_{M^4} d^4x\Biggl[\sum_{m_n\ne 0}\bar{\psi}_{4,n}(i\gamma^{\mu}D_{\mu}^{(0)}-m_n)\psi_{4,n} \notag \\
&\hspace{30mm}+
i\bar{\psi}_{R,0}\gamma^{\mu}D_{\mu}^{(0)}\psi_{R,0}+i\bar{\psi}_{L,0}\gamma^{\mu}D_{\mu}^{(0)}\psi_{L,0}
 \notag \\
&\hspace{50mm}+\left(A_{M}^{(n)}(x)\mathrm{-dependent~terms}\right)\Biggr],
\end{align} 
where $D_{\mu}^{(0)}\equiv \partial_{\mu}-iqA_{\mu}^{(0)}(x)$ with a
zero mode of gauge field $A_{\mu}^{(0)}(x)$, and $A_{M}^{(n)}(x)$ are
higher modes.
Here, we used $\psi_{4,n}\equiv \psi_{4R,n}+\psi_{4L,n}$ and the mass equations with background  gauge fields $A_{z}^{(b)}(z)$ and $A_{\bar{z}}^{(b)}(z)$,
\begin{align}
&-2D_z^{(b)}\psi_{-,n}(z) =m_n\psi_{+,n}(z), ~~~2D_{\bar{z}}^{(b)}\psi_{+,n}(z) =m_n\psi_{-,n}(z), \notag \\
&D_z^{(b)}\equiv \partial_z -iqA_z^{(b)}(z),~~~D_{\bar{z}}^{(b)}\equiv \partial_{\bar{z}} -iqA_{\bar{z}}^{(b)}(z).
\label{mass_equation}
\end{align}
We have also used the orthonormality condition 
\begin{align}
\int_{T^2}dzd\bar{z}~\psi_{\pm,n}(z)\psi_{\pm,m}^*(z)=\delta_{nm}.
\end{align}

Our first interest is the feature of zero modes for a fermion with $m_n=0$, and its interaction with magnetic flux.
We will investigate the zero-mode parts of the two-dimensional Weyl
fermion $\psi_{2\pm,n}(z)$ and the background  gauge fields 
$A_{z}^{(b)}(z)$ and $A_{\bar{z}}^{(b)}(z)$.

\subsection{Magnetic flux quantization on $T^2$}

We review the $U(1)$ gauge theory on a two-dimensional torus 
with magnetic flux following ref.\cite{Hashimoto:1997gm,Cremades:2004wa}.
The complex coordinate $z$ on one-dimensional complex plane satisfies
the identification $z\sim z+1\sim z+\tau~(\tau \in
\mathbb{C},\mathrm{Im}\tau > 0)$ 
on $T^2$.\footnote
{For convenience, we choose $(1,\tau)$ as two circumferences of the two-dimensional torus.}
The non-zero magnetic flux $b$ on $T^2$ can be obtained as $b=\int_{T^2}F^{(b)}$ with the field strength
\begin{align}
F^{(b)}={ib\over 2\mathrm{Im}\tau}dz\wedge d\bar{z} \equiv F_{z\bar{z}}^{(b)}dzd\bar{z}.
\label{Fb}
\end{align} 
For $F^{(b)}=dA^{(b)}$, the vector potential $A^{(b)}$ can be written as
\begin{align}
A^{(b)}(z)
&={b\over 2\mathrm{Im}\tau}\mathrm{Im}[(\bar{z}+\bar{a}_w)dz] \notag \\
&={b\over 4i\mathrm{Im}\tau}(\bar{z}+\bar{a}_w)dz-{b\over 4i\mathrm{Im}\tau}(z+a_w)d\bar{z}  \notag \\[2mm]
&\equiv
{A_{z}^{(b)}(z)\hspace{0.5mm}dz+A_{\bar{z}}^{(b)}(z)\hspace{0.5mm}d\bar{z}
  \mathstrut} ,
\label{v_pA}
\end{align}
where $a_w$ is a complex Wilson line phase.
{}From eq.(\ref{v_pA}), we obtain 
\begin{align}
A^{(b)}(z+1)&=A^{(b)}(z)+{b\over 2\mathrm{Im}\tau}\mathrm{Im}dz 
\equiv A^{(b)}(z)+d\chi_1 (z+a_w),~~\notag  \\
A^{(b)}(z+\tau )&=A^{(b)}(z)+{b\over 2\mathrm{Im}\tau}\mathrm{Im}(\bar{\tau}dz) 
\equiv A^{(b)}(z)+d\chi_{\tau} (z+a_w),
\label{v_pAtra}
\end{align}
where $\chi_1(z+a_w)$ and $\chi_{\tau}(z+a_w)$ are given by\footnote{
Note that we can freely add constants in the definition of $\chi_1$ and $\chi_{\tau}$ without changing the relation (\ref{v_pAtra}).
Here, 
we chose such constants, as in eq.(\ref{def of chi}), for later convenience.}
\begin{align}
\chi_1(z+a_w)={b\over 2\mathrm{Im}\tau}\mathrm{Im}(z+a_w),~~
\chi_{\tau}(z+a_w)={b\over 2\mathrm{Im}\tau}\mathrm{Im}[\bar{\tau}(z+a_w)].
\label{def of chi}
\end{align}

Moreover, we require the Lagrangian density $\mathcal{L}_{\mathrm{6D}}$ (\ref{6DLagrangian}) to be single-valued, i.e.,
\begin{align}
\mathcal{L}_{\mathrm{6D}}(A(x,z),\Psi_+(x,z)) 
&=\mathcal{L}_{\mathrm{6D}}(A(x,z+1),\Psi_+(x,z+1)) \notag \\
&=\mathcal{L}_{\mathrm{6D}}(A(x,z+\tau),\Psi_+(x,z+\tau)).
\label{LPhi_T^2}
\end{align}
Then, this field $\Psi_+(x,z)$ should satisfy the pseudo periodic boundary conditions
\begin{align}
\Psi_+(x,z+1) =U_1(z)\Psi_+(x,z),~~
\Psi_+(x,z+\tau) =U_{\tau}(z)\Psi_+(x,z), 
\label{Psi_M^4T^2}
\end{align}
i.e.,
\begin{align}
\psi_{\pm,n}(z+1) =U_1(z)\psi_{\pm,n}(z),~~
\psi_{\pm,n}(z+\tau) =U_{\tau}(z)\psi_{\pm,n}(z), 
\label{psi_T^2}
\end{align}
with
\begin{align}
U_1(z)\equiv e^{iq\chi_1(z+a_w) +2\pi i \alpha_1}, ~~~U_{\tau}(z)\equiv e^{iq\chi_{\tau}(z+a_w) +2\pi i \alpha_{\tau}},
\label{transformations for lattice shifts}
\end{align}
where $\alpha_1$ and $\alpha_{\tau}$ are allowed to be any real number, and are called Scherk-Schwarz phases.
The consistency of the contractible loops, e.g., $z\to z+1\to z+1+\tau\to z+\tau\to z$, requires the magnetic flux quantization condition,
\begin{align}
{qb \over 2\pi}\equiv M\in \mathbb{Z}.
\end{align}
Then, $U_1(z)$ and $U_{\tau}(z)$ satisfy 
\begin{align}
U_1(z+\tau)U_{\tau}(z) =U_{\tau}(z+1)U_1(z).
\end{align}

It should be emphasized that all of the Wilson line phase and the
Scherk-Schwarz phases can be arbitrary, but are not physically
independent because the Wilson line phase can be absorbed into the
Scherk-Schwarz phases by a redefinition of fields and 
vice versa (see appendix \ref{Wlp&alp}).
This fact implies that we can take, for instance, the basis of vanishing Wilson line phases, without any loss of generality.
It is then interesting to point out that allowed Scherk-Schwarz phases
are severely restricted for $T^2/Z_N$ orbifold models, as we will see
in the next section, 
while there is no restriction on the Scherk-Schwarz phases for $T^2$ models.

\subsection{Zero-mode solutions of a fermion}

We focus on zero-mode solutions $\psi_{\pm,0}(z)$ with $m_n=0$ on $T^2$ with magnetic flux.
{}From eqs.(\ref{mass_equation}) and (\ref{v_pA}), $\psi_{\pm,0}(z)$ satisfy the zero-mode equations 
\begin{align}
\left(\partial_{\bar{z}} +{\pi M \over 2\mathrm{Im}\tau}(z+a_w)\right)\psi_{+,0} (z;a_w)=0,~~
\left(\partial_z -{\pi M \over 2\mathrm{Im}\tau}(\bar{z}+\bar{a}_w)\right)\psi_{-,0} (z;a_w)=0.
\label{psi_pm}
\end{align}
Here and hereafter, to emphasize the existence of the Wilson line phase, we will rewrite $\psi_{\pm,0}(z)$ as $\psi_{\pm,0}(z;a_w)$.
The fields $\psi_{\pm,0}(z;a_w)$ should obey the conditions (\ref{psi_T^2}), i.e.,
\begin{align}
&\psi_{\pm,0} (z+1;a_w) =e^{iq\chi_1(z+a_w) +2\pi i \alpha_1}\psi_{\pm,0}(z;a_w),~~\notag \\
&\psi_{\pm,0}(z+\tau ;a_w)=e^{iq\chi_{\tau}(z+a_w) +2\pi i \alpha_{\tau}}\psi_{\pm,0} (z;a_w).
\label{psi_trans1tau}
\end{align}
Then, the zero-mode solutions of $\psi_{\pm,0}(z;a_w)$ with the Wilson line phase are found to be of the form  
\begin{align}
\psi_{+,0}(z;a_w)
&=\mathcal{N} e^{i\pi M (z+a_w){\mathrm{Im}(z+a_w)\over \mathrm{Im}\tau}}\cdot \vartheta \left[
\begin{array}{c}
{j+\alpha_1 \over M} \\ -\alpha_{\tau}
\end{array}
\right] (M(z+a_w),M\tau) \notag \\
&\equiv \psi_{+,0}^{(j+\alpha_1,\alpha_{\tau})} (z;a_w)~~~~~{\rm for}~ M>0,
\label{psi_p_func} \\
\psi_{-,0}(z;a_w)
&=\mathcal{N} e^{i\pi M (\bar{z}+\bar{a}_w){\mathrm{Im}(\bar{z}+\bar{a}_w)\over \mathrm{Im}\bar{\tau}}}\cdot \vartheta \left[
\begin{array}{c}
{j+\alpha_1 \over M} \\ -\alpha_{\tau}
\end{array}
\right] (M(\bar{z}+\bar{a}_w),M\bar{\tau}) \notag \\
&\equiv \psi_{-,0}^{(j+\alpha_1,\alpha_{\tau})} (z;a_w) ~~~~~ {\rm for}~ M<0,
\label{psi_m_func}
\end{align}
where $j=0,1,\cdots, |M|-1$, and  $\mathcal{N}$ is the normalization factor.
For $a_w=0$ and $(\alpha_1,\alpha_{\tau})=(0,0)$,
$\psi_{\pm,0}^{(j+\alpha_1,\alpha_{\tau})}(z;a_w)$ are reduced to the
results obtained 
in ref.\cite{Cremades:2004wa}.
We would like to note the two features that for $M>0~(M<0)$, only $\psi_{+,0}~(\psi_{-,0})$ has solutions, and that the number of solutions is given by $|M|$.
Thus, we can obtain a $|M|$-generation chiral theory in four-dimensional space-time from eq.(\ref{6DLagrangian}). 
Here, $\mathcal{N}$ may be fixed by the orthonormality condition
\begin{align}
\int_{T^2}dzd\bar{z}~\psi_{\pm,0}^{(j+\alpha_1,\alpha_{\tau})} (z;a_w)(\psi_{\pm,0}^{(k+\alpha_1,\alpha_{\tau})} (z;a_w))^*=\delta^{jk},
\end{align}
and is given by $\mathcal{N}=\left(2\mathrm{Im}\tau |M| \over \mathcal{A}^2\right)^{1\over 4}$ with the area of the torus $\mathcal{A}$.

The $\vartheta$ function is defined by 
\begin{align}
&\vartheta \left[
\begin{array}{c}
a\\ b
\end{array}
\right] (c\nu,c\tau) 
=\sum_{l=-\infty}^{\infty}e^{i\pi (a+l)^2c\tau}e^{2\pi i(a+l)(c\nu +b)} ,
\end{align}
with the properties
\begin{align}
&\vartheta \left[
\begin{array}{c}
a\\ b
\end{array}
\right] (c(\nu +n),c\tau)
=e^{2\pi i acn}\vartheta \left[
\begin{array}{c}
a\\ b
\end{array}
\right] (c\nu,c\tau), \notag \\
&\vartheta \left[
\begin{array}{c}
a\\ b
\end{array}
\right] (c(\nu +n\tau),c\tau)
=e^{-i\pi cn^2\tau -2\pi i n(c\nu +b)}\vartheta \left[
\begin{array}{c}
a\\ b
\end{array}
\right] (c\nu,c\tau), \notag \\
&\vartheta \left[
\begin{array}{c}
a+m\\ b+n
\end{array}
\right] (c\nu,c\tau)
=e^{2\pi i an}\vartheta \left[
\begin{array}{c}
a\\ b
\end{array}
\right] (c\nu,c\tau), 
\end{align}
where $a$ and $b$ are real numbers, $c$, $m$ and $n$ are integers, and $\nu$ and $\tau$ are complex numbers with $\mathrm{Im}\tau >0$.

\section{Twisted orbifolds with magnetic flux \label{sec:Twi_orbi w/mf}}

In the previous section, we reviewed the $U(1)$ gauge theory on a two-dimensional torus $T^2$ with magnetic flux.
Then, we found that the number of zero-mode fermions is given by the magnitude of magnetic flux $|M|$.
In this section, we study the $U(1)$ gauge theory on twisted orbifolds
$T^2/Z_N$ with magnetic flux, and investigate the degeneracy of
zero-mode solutions and the allowed values of the Wilson line phase
$a_w$ 
and the Scherk-Schwarz phases $\alpha_1$ and $\alpha_{\tau}$.

\subsection{$T^2/Z_N$ twisted orbifold}

A two-dimensional twisted orbifold $T^2/Z_N$ is defined by dividing a
one-dimensional complex plane by lattice shifts $t_1$, $t_{\tau}$ and
a $Z_N$ discrete rotation (twist) $s$ 
such as
\begin{align}
t_1:z \to z+1,~~~t_{\tau}:z \to z+\tau ,~~~s:z \to \omega z,
\label{translation_T2/ZN}
\end{align}
with
\begin{align}
\omega\equiv e^{2\pi i/{N}}.
\end{align}
Thus, the orbifold obeys the identification 
\begin{align}
z\sim \omega z +m+n\tau ~~~~~\mathrm{for}~~{}^{\forall} m,n \in \mathbb{Z}.
\label{identification_z}
\end{align}
It has already been known that there exist only four kinds of the orbifolds such as $T^2/Z_N~(N=2,3,4,6)$.
We would like to note the relation between the moduli $\tau$ and the rotation $\omega$ for each orbifold.
For $N=2$, there is no limitation on $\tau$ except for
$\mathrm{Im}\tau >0$, but for $N=3,4,6$, $\tau$ should be equivalent to
$\omega$ because of the analysis 
by crystallography \cite{Choi:2006qh}.
For convenience, we still use both $\tau$ and $\omega$ as a base
vector on the lattice and the $Z_N$ twist, respectively below, though
$\tau =\omega$ 
for $N=3,4,6$.

Moreover, an important feature is the existence of fixed points $z_{\mathrm{fp}}$ defined by
\begin{align}
z_{\mathrm{fp}}=\omega z_{\mathrm{fp}} + m+n\tau ~~~~~\mathrm{for}~~{}^{\exists}m,n\in \mathbb{Z}.
\label{fixed pints}
\end{align} 
Since each fixed point is specified by the $Z_N$ twist $\omega$ and
the 
shift $m+n\tau$, we define $z_{\mathrm{fp}}$ as $(\omega, m+n\tau)$ 
with the language of space group.
On the complex plane, there exist an infinite number of fixed points because the possible combinations of $(m,n)$ exist countlessly.
On the torus, however, $z_{\mathrm{fp}}$ and
$z_{\mathrm{fp}}+m'+n'\tau~({}^{\forall} m',n'\in \mathbb{Z})$ should
be identified with the torus identification 
$z\sim z+m+n\tau$.
Then, it follows from eq.(\ref{fixed pints}) that since $z_{\mathrm{fp}}+m'+n'\tau$ satisfies the relation
\begin{align}
z_{\mathrm{fp}}+m'+n'\tau =\omega (z_{\mathrm{fp}}+m'+n'\tau) +m+n\tau,
\end{align}
i.e.,
\begin{align}
z_{\mathrm{fp}}=\omega z_{\mathrm{fp}}+(\omega -1)(m'+n'\tau) +m+n\tau,
\end{align}
$(\omega, m+n\tau)$ should be identified with $(\omega,(\omega -1)(m'+n'\tau) +m+n\tau)$.

For example, let us consider the case of $T^2/Z_3$.
For $n'=0$, $(\omega, m+n\tau)$ is identified with $(\omega, m-m'+(n+m')\tau)$.
For $m'=2n'\ne 0$, $(\omega, m+n\tau)$ is also identified with $(\omega, m-3n'+n\tau)$.
{}From these identifications, one can find three fixed points, i.e.,
\begin{align}
(m,n)\equiv (0,0),(1,0),(2,0)~~~~~\mathrm{mod}~~(-1,1)~\mathrm{and}~(3,0).
\end{align}
Actually, the three fixed points on the fundamental region are given by
\begin{align}
z_{\mathrm{fp}}=0,~{2+\tau \over 3},~{1+2\tau \over 3}~~~~~\mathrm{for}~~(m,n)=(0,0),(1,0),(1,1).
\end{align}

In the same way,\footnote{
See in detail ref.\cite{Kobayashi:1991rp}.} 
on $T^2/Z_2$, four fixed points exist, which are
$z_{\mathrm{fp}}=0,{1\over 2},{\tau \over 2},{1+\tau \over 2}$ for
$(m,n)=(0,0),(1,0),(0,1),(1,1)$, 
respectively.
On $T^2/Z_4$, two fixed points exist, which are $z_{\mathrm{fp}}=0,{1+\tau \over 2}$ for $(m,n)=(0,0),(1,0)$, respectively.
On $T^2/Z_6$, only one fixed point exists, which is $z_{\mathrm{fp}}=0$ for $(m,n)=(0,0)$.
The fundamental region and the fixed points for each orbifold are depicted in figure~\ref{fig:one}.
As we will see below, the number of fixed points correspond to the variety of allowed Scherk-Schwarz phases on each orbifold.

\begin{figure}[t]
\begin{center}
\hspace{8mm}
\includegraphics[width=100mm]{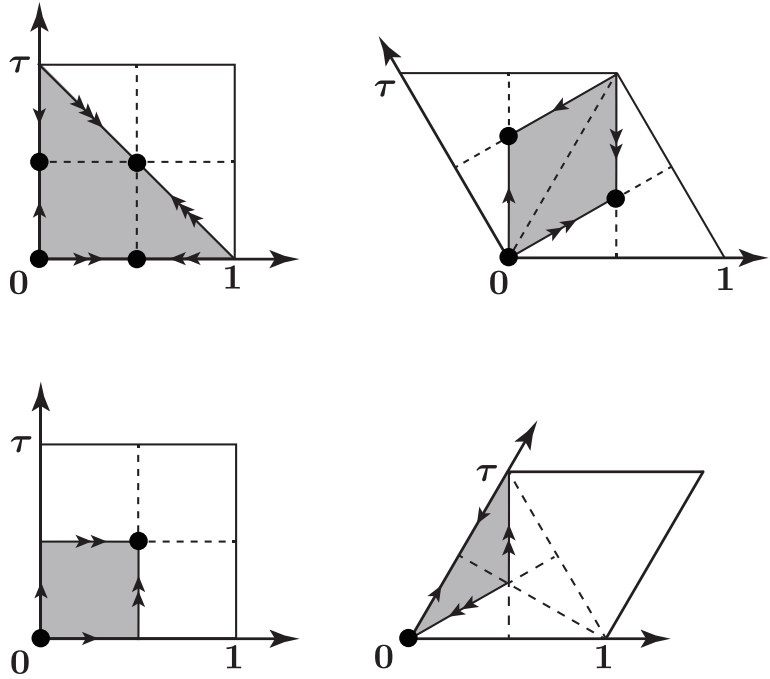}
\end{center}
\caption{The fundamental regions of $T^2/Z_N$ orbifolds $(N=2,3,4,6)$ and their fixed points. 
The top left is $T^2/Z_2$, the top right is $T^2/Z_3$, the down left is $T^2/Z_4$, and the down right is $T^2/Z_6$. 
The black dots denote the fixed points, and the gray areas correspond to the fundamental region. 
Each of single, double and triple arrows means the direction and pair of identification by lattice shifts and $Z_N$ twist.}
\label{fig:one}
\end{figure}

\subsection{Field theory on orbifold\label{sec:Field theory on orbifold}}

Next, we study a field theory on the orbifold. 
Let us consider the following Lagrangian density on six-dimensional space-time with the orbifold $T^2/Z_N$,
\begin{align}
\mathcal{L}_{\mathrm{6D}}^{\mathrm{Weyl}}\equiv i\bar{\Psi}_{T^2/Z_N+}(x,z)\Gamma^M(\partial_M -iq A_M(x,z))\Psi_{T^2/Z_N+}(x,z),
\end{align}
where $\Psi_{T^2/Z_N+}(x,z)$ is a six-dimensional Weyl fermion on $M^4\times T^2/Z_N$.
Here and hereafter, we take the gauge with $a_w=0$ because the Wilson line phase can be absorbed into the Scherk-Schwarz phases (see appendix \ref{Wlp&alp}).
In order to define a field theory on the orbifold, we need to specify
what are boundary conditions under the lattice shifts $t_1$,
$t_{\tau}$ and the $Z_N$ twist $s$ 
for the fermion.
Then, we define the boundary conditions for $\Psi_{T^2/Z_N+}(x,z)$ as 
\begin{align}
&\Psi_{T^2/Z_N+}(x,z+1)=U_1(z)\Psi_{T^2/Z_N+}(x,z), \notag \\
&\Psi_{T^2/Z_N+}(x,z+\tau)=U_{\tau}(z)\Psi_{T^2/Z_N+}(x,z), \notag \\
&\Psi_{T^2/Z_N+}(x,\omega z)= \mathcal{S}V(z)\Psi_{T^2/Z_N+}(x,z) ,
\label{general_psi_BC}
\end{align}
where $\mathcal{S}\equiv
(\Gamma^{z}\Gamma^{\bar{z}}+\omega\Gamma^{\bar{z}}\Gamma^{z})/4=\mathrm{diag}(\mathbf{1}_{4\times
  4},\omega \mathbf{1}_{4\times 4})$, 
and $V(z)$ is a transformation function for the $Z_N$ twist.
The transformation functions $U_1(z)$ and $U_{\tau}(z)$ for the
lattice shifts $t_1$ and $t_{\tau}$ are given in
eq.(\ref{transformations for lattice shifts}) 
with $a_w=0$.
However, the Scherk-Schwarz phases $\alpha_1$ and $\alpha_{\tau}$
cannot be freely chosen, and are allowed to be certain discrete values
on the orbifold, 
as we will see later.

In a way similar to eq.(\ref{6D_Dirac to two Weyl}), we can expand the Weyl fermion $\Psi_{T^2/Z_N+}(x,z)$ on $M^4\times T^2/Z_N$ such as
\begin{align}
\Psi_{T^2/Z_N+}(x,z)&=\psi_{T^2/Z_N R}(x,z) +\psi_{T^2/Z_N L}(x,z) \notag \\
&=\sum_n\left(\psi_{4R,n}(x) \otimes \psi_{T^2/Z_N2+,n}(z) + \psi_{4L,n}(x) \otimes \psi_{T^2/Z_N2-,n}(z) \right), \label{6D_Dirac to two Weyl on T^2/Z_N}
\end{align}
with
\begin{align}
\psi_{T^2/Z_N2+,n}(z)=\left(
\begin{array}{c}
\psi_{T^2/Z_N+,n} (z)\\ 0
\end{array}
\right), ~~~
\psi_{T^2/Z_N2-,n}(z)=\left(
\begin{array}{c}
0\\ \psi_{T^2/Z_N-,n}(z)
\end{array}
\right).
\end{align}
Then, the boundary conditions (\ref{general_psi_BC}) for $\Psi_{T^2/Z_N+}(x,z)$ are replaced by those for $\psi_{T^2/Z_N\pm,n}(z)$, i.e.,
\begin{align}
&\psi_{T^2/Z_N\pm,n}(z+1)=U_1(z)\psi_{T^2/Z_N\pm,n}(z), \notag \\
&\psi_{T^2/Z_N\pm,n}(z+\tau) =U_{\tau}(z)\psi_{T^2/Z_N\pm,n}(z), \notag \\
&\psi_{T^2/Z_N+,n}(\omega z)= V(z)\psi_{T^2/Z_N+,n}(z), \notag \\
&\psi_{T^2/Z_N-,n}(\omega z)= \omega V(z)\psi_{T^2/Z_N-,n}(z).
\label{general_psi_BC_extra}
\end{align}
Here, it is worthwhile to note that the wave functions
$\psi_{T^2/Z_N\pm,n}(z)$ on the orbifold $T^2/Z_N$ can be constructed
from certain linear combinations 
of $\psi_{\pm,n}(z)$ on the torus $T^2$.
This is because the orbifold $T^2/Z_N$ is obtained by dividing the torus $T^2$ by the $Z_N$ discrete rotation.

{}From eq.(\ref{v_pA}) with $a_w=0$, $A_{z}^{(b)}(z)$ and $A_{\bar{z}}^{(b)}(z)$ satisfy 
\begin{align}
A_{z}^{(b)}(\omega z)=\bar{\omega}A_{z}^{(b)}(z),~~~A_{\bar{z}}^{(b)}(\omega z)=\omega A_{\bar{z}}^{(b)}(z).
\end{align}
The transformation function $V(z)$ is given by
\begin{align}
V(z)=e^{2\pi i \beta},
\label{V(x,z)}
\end{align}  
where $\beta$ is a real number.
{}From the requirement that $N$ times the twisted transformation in eq.(\ref{translation_T2/ZN}) should be identical to the identity operation, i.e., $s^N=1$, $\beta$ has to satisfy 
\begin{align}
\beta N\equiv 0~~~~~\mathrm{mod}~~1.
\end{align}

When we require that $\mathcal{L}_{\mathrm{6D}}^{\mathrm{Weyl}}$ is
single-valued under the lattice shifts and the $Z_N$ twist, the
boundary conditions 
for the gauge fields $A_M(x,z)$ can be obtained as 
\begin{align}
&A_M(x,z+1)=U_1(z)\left(A_M(x,z) -{i\over q}\partial_M\right) U_1^{\dag}(z), \notag \\
&A_M(x,z+\tau)=U_{\tau}(z)\left(A_M(x,z) -{i\over q}\partial_M\right) U_{\tau}^{\dag}(z), \notag \\
&A_{\mu}(x,\omega z)=A_{\mu}(x,z),~~~A_{z}(x,\omega z)=\bar{\omega}A_{z}(x,z), \notag \\
&A_{\bar{z}}(x,\omega z)=\omega A_{\bar{z}}(x,z).
\end{align}

Moreover, let us investigate the boundary conditions for general lattice shifts $m+n\tau ~(m,n\in \mathbb{Z})$ and $Z_N$ twists $\omega^k~(k\in \mathbb{Z})$.
To this end, we define the transformation function $U_{m+n\tau}(z)$ through the relation
\begin{align}
\Psi_{T^2/Z_N}(x,z+m+n\tau) =U_{m+n\tau}(z)\Psi_{T^2/Z_N}(x,z).
\label{def of U_m+ntau}
\end{align}
Then, we obtain
\begin{align}
\Psi_{T^2/Z_N}(x,\omega^k (z+m+n\tau)) =U_{\omega^k( m+n\tau)}(\omega^k z)\Psi_{T^2/Z_N}(x,\omega^k z),
\end{align}
because $\omega^k(m+n\tau)$ for ${}^{\forall}k,m,n\in \mathbb{Z}$ can
be equivalently expressed as a lattice shift $m'+n'\tau$ 
for ${}^{\exists} m',n'\in \mathbb{Z}$. 

{}From the above definition (\ref{def of U_m+ntau}), $U_{m+n\tau}(z)$ turns out to satisfy 
\begin{align}
&U_{m+1+n\tau}(z)=U_1(z+n\tau) U_{m+n\tau}(z), \notag \\
&U_{m+(n+1)\tau}(z)=U_{\tau}(z+m) U_{m+n\tau}(z), \notag \\
&U_{-(m+n\tau)}(z)=U_{m+n\tau}^{\dag}(z-m-n\tau)=U_{m+n\tau}^{-1}(z-m-n\tau), \notag \\
&U_{m}(z)=(U_1(z))^m,~~~U_{n\tau}(z)=(U_{\tau}(z))^n,
\label{relations of U_m+ntau}
\end{align}
where we have used the relations $U_1(z+m)=U_1(z)$ and
$U_{\tau}(z+n\tau)=U_{\tau}(z)$, which will be derived from
eqs.(\ref{def of chi}) and 
(\ref{transformations for lattice shifts}).
We can further show that from eq.(\ref{general_psi_BC}), $U_{m+n\tau}(z)$ should obey the relation
\begin{align}
U_{m+n\tau}(z)=U_{\omega^k(m+n\tau)}(\omega^k z)~~~~~\mathrm{for}~~k\in \mathbb{Z}.
\label{Umnt-Uo(mnt)_iden}
\end{align}
It follows that we find 
\begin{align}
&U_1(z)=U_{-1}(-z),~~~U_{\tau}(z)=U_{-\tau}(-z) \hspace{13mm} \mathrm{for}~~N=2, \notag \\
&U_1(z)=U_{\omega}(\omega z),~~~U_{\tau}(z)=U_{\omega \tau}(\omega z) \hspace{16.6mm} \mathrm{for}~~N=3,4,6.
\label{FE_U1Utau_aw0}
\end{align}

\subsubsection{Scherk-Schwarz phases without magnetic flux}

First, let us review Scherk-Schwarz phases without magnetic flux \cite{Kobayashi:1991rp,Kobayashi:1990mi}.
Then, $U_1(z)$ and $U_{\tau}(z)$ are independent of $z$,
\begin{align}
U_1(z)=e^{2\pi i\alpha_1},~~~U_{\tau}(z)=e^{2\pi i \alpha_{\tau}}.
\end{align}
For $N=2$, using eqs.(\ref{relations of U_m+ntau}) and (\ref{FE_U1Utau_aw0}), we obtain
\begin{align}
e^{2\pi i \alpha_1} = e^{-2\pi i \alpha_1}, ~~~ 
e^{2\pi i \alpha_\tau} = e^{-2 \pi i\alpha_\tau},
\end{align}
i.e.,
\begin{align}
(\alpha_1,\alpha_\tau)\equiv (0,0), ~ (1/2,0),  ~  (0,1/2), ~  (1/2,1/2),  ~~~~~\mathrm{mod}~~1.
\label{SSphase_T2/Z2w/omf}
\end{align}

For $N=3$, using eqs.(\ref{relations of U_m+ntau}) and (\ref{FE_U1Utau_aw0}), we obtain
\begin{align}
e^{2\pi i \alpha_1} = e^{2\pi i \alpha_\tau}, ~~~ 
e^{2\pi i \alpha_\tau} = e^{-2 \pi i(\alpha_1 + \alpha_\tau )},
\end{align}
i.e.,
\begin{align}
(\alpha_1,\alpha_\tau)\equiv (0,0), ~  (1/3,1/3), ~ (2/3,2/3), ~~~~~\mathrm{mod}~~1,
\label{SSphase_T2/Z3w/omf}
\end{align}
where we have used the relations that $\omega \tau =-1-\tau$ for $\tau=\omega=e^{2\pi i/3}$.

For $N=4$, using eqs.(\ref{relations of U_m+ntau}) and (\ref{FE_U1Utau_aw0}), we obtain
\begin{align}
e^{2\pi i \alpha_1} = e^{2\pi i \alpha_\tau}, ~~~ 
e^{2\pi i \alpha_\tau} = e^{-2 \pi i\alpha_1 },
\end{align}
i.e.,
\begin{align}
(\alpha_1,\alpha_\tau)\equiv (0,0), ~ (1/2,1/2),  ~~~~~\mathrm{mod}~~1,
\label{SSphase_T2/Z4w/omf}
\end{align}
where we have used the relations that $\omega \tau =-1$ for $\tau=\omega=i$.

For $N=6$, using eqs.(\ref{relations of U_m+ntau}) and (\ref{FE_U1Utau_aw0}), we obtain
\begin{align}
e^{2\pi i \alpha_1} = e^{2\pi i \alpha_\tau}, ~~~ 
e^{2\pi i \alpha_\tau} = e^{2 \pi i(-\alpha_1+\alpha_\tau) },
\end{align}
i.e.,
\begin{align}
(\alpha_1,\alpha_\tau)\equiv (0,0), ~~~~~\mathrm{mod}~~1,
\label{SSphase_T2/Z6w/omf}
\end{align}
where we have used the relations that $\omega \tau =-1+\tau$ for $\tau=\omega=e^{\pi i/3}$.

Here, we would like to note that the variety of the Scherk-Schwarz
phases $(\alpha_1,\alpha_{\tau})$ corresponds to the number of fixed
points in the fundamental region on each orbifold.

\subsubsection{Scherk-Schwarz phases with magnetic flux}

Next, let us investigate the Scherk-Schwarz phases with magnetic flux.
Then, $U_1(z)$ and $U_{\tau}(z)$ depend on $z$,
\begin{align}
U_1(z)=e^{iq\chi_1(z) +2\pi i\alpha_1},~~~U_{\tau}(z)=e^{iq\chi_{\tau}(z) +2\pi i \alpha_{\tau}}.
\end{align}
We apply eqs. (\ref{relations of U_m+ntau}) and (\ref{FE_U1Utau_aw0})
to these $U_1(z)$ and $U_\tau(z)$.
Then, for $N=2,4$, we have the same results as
eqs.(\ref{SSphase_T2/Z2w/omf}) and (\ref{SSphase_T2/Z4w/omf}),
respectively.
On the other hand, 
we will show that the allowed Scherk-Schwarz phases with magnetic flux
are different from those obtained in the previous section 
for $N=3,6$ with $M=\mathrm{odd}$. 

For $N=3$, using eqs.(\ref{relations of U_m+ntau}) and (\ref{FE_U1Utau_aw0}), we obtain
\begin{align}
e^{2\pi i \alpha_1} = e^{2\pi i \alpha_\tau}, ~~~ 
e^{2\pi i \alpha_\tau} = e^{-2 \pi i(\alpha_1 + \alpha_\tau )+i\pi M},
\end{align}
i.e.,
\begin{align}
(\alpha_1,\alpha_\tau) &=(0, 0),~(1/3,1/3),~(2/3,2/3) \hspace{20mm} \mathrm{for}~~M=\mathrm{even}, \notag \\
(\alpha_1,\alpha_\tau) &=(1/6,1/6),~(1/2,1/2),~(5/6,5/6) \hspace{12mm} \mathrm{for}~~M=\mathrm{odd}.
\label{SSphase_T2/Z3w/mf}
\end{align}

For $N=6$, using eqs.(\ref{relations of U_m+ntau}) and (\ref{FE_U1Utau_aw0}), we obtain
\begin{align}
e^{2\pi i \alpha_1} = e^{2\pi i \alpha_\tau}, ~~~ 
e^{2\pi i \alpha_\tau} = e^{2 \pi i(-\alpha_1 + \alpha_\tau )+i\pi M},
\end{align}
i.e.,
\begin{align}
(\alpha_1,\alpha_\tau) &=(0, 0), \hspace{20mm} \mathrm{for}~~M=\mathrm{even}, \notag \\
(\alpha_1,\alpha_\tau) &=(1/2,1/2),\hspace{12mm} \mathrm{for}~~M=\mathrm{odd}.
\label{SSphase_T2/Z6w/mf}
\end{align}

The allowed Scherk-Schwarz phases are shown in table~\ref{table:SS phases}.
It is found that the variety of the Scherk-Schwarz phases still
corresponds to the number of fixed points even with non-zero magnetic
flux.\footnote{Also, the same relation among the fixed points,
  discrete Wilson lines as well as phases is studied in 
intersecting D-brane models on $T^6/(Z_2 \times Z_2)$\cite{Blumenhagen:2005tn}.}
However, it is remarkable that the non-zero magnetic flux with
$M=\mathrm{odd}$ affects the values of the Scherk-Schwarz phases for $N=3,6$, and especially does not permit them to vanish. 

\begin{table}[htbp]
\vspace{3mm}
\begin{center}
\begin{tabular}{|c|c|c|}
\hline
Orbifold &$M$& Scherk-Schwarz phases $(\alpha_1,\alpha_{\tau})~(\mathrm{mod}~1)$ \\
\hline
$T^2/Z_2$&-&$(0,0), ~ ({1\over 2},0),  ~  (0,{1\over 2}), ~  ({1\over 2},{1\over 2})$ \\ \hline
\multirow{2}{*}{$T^2/Z_3$}&even&$(0, 0),~({1\over 3},{1\over 3}),~({2\over 3},{2\over 3})$ \\ \cline{2-3}
&odd&$ ({1\over 6},{1\over 6}),~ ({1\over 2},{1\over 2}),~({5\over 6},{5\over 6})$ \\ \hline
$T^2/Z_4$&-&$(0,0), ~   ({1\over 2},{1\over 2})$ \\ \hline
\multirow{2}{*}{$T^2/Z_6$}&even&$(0,0)$ \\ \cline{2-3}
&odd&$ ({1\over 2},{1\over 2})$ \\
\hline
\end{tabular}
\end{center}
\vspace{3mm}
\caption{The list of the allowed Scherk-Schwarz phases with magnetic flux.}
\label{table:SS phases}
\end{table}

\subsection{$Z_N$ eigenstates of fermions\label{sec:Z_Neigenstate}}

Here, we explain how to construct the wave functions $\psi_{T^2/Z_N\pm,n}(z)$ on $T^2/Z_N$ from the wave functions $\psi_{\pm,n}(z)$ on $T^2$.
To clearly distinguish wave functions on $T^2$ from those on $T^2/Z_N$, we rewrite the wave functions with $a_w=0$ on $T^2$ as $\psi_{T^2\pm,n}(z)$.
Since the wave functions $\psi_{T^2\pm,n}(z)$ should obey the desired
boundary conditions (\ref{psi_T^2}) as well as the zero-mode wave
functions
$\psi_{T^2\pm,0}^{(j+\alpha_1,\alpha_{\tau})}(z)=\psi_{\pm,0}^{(j+\alpha_1,\alpha_{\tau})}(z;0)$,
we rewrite them as
$\psi_{T^2\pm,n}^{(j+\alpha_1,\alpha_{\tau})}(z)~(j=0,1,\cdots
,|M|-1)$, which are constructed by a way similar to the analysis of
harmonic oscillator 
in the quantum mechanics (see section~\ref{sec:KK and masses}).   

In addition to the torus boundary conditions (the first two conditions
of eq.(\ref{general_psi_BC_extra})), the wave functions
$\psi_{T^2/Z_N\pm,n}(z)$ have to satisfy the orbifold boundary
conditions 
(the last two conditions of eq.(\ref{general_psi_BC_extra}))
\begin{align}
&\psi_{T^2/Z_N+,n}(\omega z)= \eta \psi_{T^2/Z_N+,n}(z), \notag \\
&\psi_{T^2/Z_N-,n}(\omega z)= \omega \eta \psi_{T^2/Z_N-,n}(z),
\end{align}
with
\begin{align}
\eta\equiv e^{2\pi i \beta},
\end{align}
i.e.,
\begin{align}
\eta \in \{1,\omega,\omega^2,\cdots, \omega^{N-1} \}.
\end{align}

Then, we can construct $\psi_{T^2/Z_N\pm,n}(z)$ by the following linear combinations of $\psi_{T^2\pm,n}^{(j+\alpha_1,\alpha_{\tau})}(z)$,
\begin{align}
\psi_{T^2/Z_N+,n}^{(j+\alpha_1,\alpha_{\tau})}(z)
&=\mathcal{N}^{(j)}_{+,\eta}\sum_{k =0}^{N-1}\bar{\eta}^{k} \psi_{T^2+,n}^{(j+\alpha_1,\alpha_{\tau})}(\omega^{k}z) 
\label{psi_T2/ZN+,n} \\
\psi_{T^2/Z_N-,n}^{(j+\alpha_1,\alpha_{\tau})}(z)
&=\mathcal{N}^{(j)}_{-,\omega \eta}\sum_{k =0}^{N-1}(\bar{\omega}\bar{\eta})^{k} \psi_{T^2-,n}^{(j+\alpha_1,\alpha_{\tau})}(\omega^{k}z) 
\label{psi_T2/ZN-,n}
\end{align}
where $j~(=0,1,\cdots ,|M|-1)$ stand for the degeneracy with respect
to the $n$-mode wave functions. 
The index of $\eta~(\omega \eta)$ for
$\psi_{T^2/Z_N+,n}^{(j+\alpha_1,\alpha_{\tau})}(z)_{\eta}~(\psi_{T^2/Z_N-,n}^{(j+\alpha_1,\alpha_{\tau})}(z)_{\omega
  \eta})$ means $Z_N$ eigenvalues on $T^2/Z_N$, and
$\mathcal{N}^{(j)}_{+,\eta}~(\mathcal{N}^{(j)}_{-,\omega \eta})$ are
normalization factors, which depend on $j$, the chirality of $\psi_{T^2\pm,n}^{(j+\alpha_1,\alpha_{\tau})}(z)$, and the $Z_N$ eigenvalues $\eta~(\omega \eta)$.
It is easy to verify that both 
$\psi_{T^2/Z_N+,n}^{(j+\alpha_1,\alpha_{\tau})}(z)_{\omega}$ and
$\psi_{T^2/Z_N-,n}^{(j+\alpha_1,\alpha_{\tau})}(z)_{\omega \eta}$ 
defined above satisfy all of the boundary conditions (\ref{general_psi_BC_extra}) on $T^2/Z_N$, as they should be.

It should be emphasized that
$\psi_{T^2/Z_N\pm,n}^{(j+\alpha_1,\alpha_{\tau})}(z)_{\omega^{\ell}}$
are mutually independent for different values of $\ell$, 
i.e., eigenvalues $\omega^\ell$ under the $Z_N$ twist, 
but are not always linearly independent for different values of $j$.
We will see this feature in the next section explicitly.

\subsection{Wilson line phase}

In section~\ref{sec:Field theory on orbifold}, we have investigated the variety of
Scherk-Schwarz phases in the gauge with $a_w=0$. 
Here, we would like to consider the case of non-zero Wilson line phases.
{}From the results given in appendix~\ref{Wlp&alp}, let us transform
the $a_w=0$ gauge into the gauge, where they satisfy 
\begin{align}
M\tilde{a}_w=\alpha_1\tau -\alpha_{\tau},~~\tilde{\alpha}_1=0,~~\tilde{\alpha}_{\tau}=0,
\label{ta1=tat=0condition}
\end{align}
where $\tilde{a}_w$ and $(\tilde{\alpha}_1,\tilde{\alpha}_{\tau})$ are the redefined Wilson line phase  and the redefined Scherk-Schwarz phases, respectively. 
Substituting the value of $(\alpha_1,\alpha_{\tau})~(\mathrm{mod}~1)$
of table~\ref{table:SS phases} into eq.(\ref{ta1=tat=0condition}), we
can obtain the allowed Wilson line phases, which are shown in table~\ref{table:Wilson line phases}.
\begin{table}[htbp]
\vspace{3mm}
\begin{center}
\begin{tabular}{|c|c|c|}
\hline
Orbifold & $M$ & Wilson line phases: $M\tilde{a}_w$ \\
\hline
$T^2/Z_2$&-&$0,~{\tau \over 2},~{1\over 2},~{1+\tau \over 2}$ \\ \hline
\multirow{2}{*}{$T^2/Z_3$}&even&$0,~{2+\tau \over 3},~{1+2\tau \over 3}$ \\ \cline{2-3}
&odd&${5+\tau \over 6},~{1+\tau \over 2},~{1+5\tau \over 6}$\\ \hline
$T^2/Z_4$&-&$0,~{1+\tau \over 2}$ \\ \hline
\multirow{2}{*}{$T^2/Z_6$}&even&$0$ \\ \cline{2-3}
&odd&${1+\tau \over 2}$ \\
\hline
\end{tabular}
\end{center}
\vspace{3mm}
\caption{The allowed Wilson line phases.}
\label{table:Wilson line phases}
\end{table}
Namely, the variety of allowed values for the Wilson line phase
corresponds to the number of 
fixed points on each orbifold.

\section{Zero-mode eigenstates on $T^2/Z_N$ \label{sec:0mode eigenstates}}

In the previous section, we have discussed the $Z_N$ eigenfunctions
$\psi_{T^2/Z_N\pm,n}^{(j+\alpha_1,\alpha_{\tau})}(z)_{\omega^{\ell}}$
on $T^2/Z_N$ and found the allowed values for the Wilson line phase
$a_w$ and the Scherk-Schwarz phases $\alpha_1$ 
and $\alpha_{\tau}$ on each orbifold.
Here, we focus on the zero-mode eigenstates for each $Z_N$ eigenvalue with $a_w=0$, and study their number for each $M$.
In particular, we will pay attention to the cases that the number of
zero-mode eigenstates is given by around three, 
because we would like to construct a three generation model.

\subsection{$T^2/Z_2$ \label{zero mode on T2/Z2}}

First, we study the case of $T^2/Z_2$ with $M>0$.\footnote{
The $U(N)$ gauge theory in $\alpha_1=\alpha_{\tau}=0$ on $T^2/Z_2$
$T^6/Z_2$ and $T^6/(Z_2\times Z'_2)$ has already been studied 
in ref.\cite{Abe:2008fi,Abe:2008sx}.} 
{}From eq.(\ref{psi_T2/ZN+,n}),  the $Z_N$ eigenstates $\psi_{T^2/Z_2+,0}^{(j+\alpha_1,\alpha_{\tau})}(z)_{\pm1}~(j=0,1,\cdots ,M-1)$ are given by 
\begin{align}
\psi_{T^2/Z_2+,0}^{(j+\alpha_1,\alpha_{\tau})}(z)_{\pm 1} =
\mathcal{N}^{(j)}_{+,\pm
  1}\left(\psi_{T^2+,0}^{(j+\alpha_1,\alpha_{\tau})}(z)\pm 
\psi_{T^2+,0}^{(j+\alpha_1,\alpha_{\tau})}(-z)\right),
\label{T2Z2_Psi_eo}
\end{align}
which satisfy the eigenvalue equations
\begin{align}
\psi_{T^2/Z_2+,0}^{(j+\alpha_1,\alpha_{\tau})}(-z)_{\pm 1}=\pm \psi_{T^2/Z_2+,0}^{(j+\alpha_1,\alpha_{\tau})}(z)_{\pm 1}.
\end{align}
{}From eq.(\ref{psi_p_func}), $\psi_{T^2+,0}^{(j+\alpha_1,\alpha_{\tau})}(z)$ possess a relation such as
\begin{align}
\psi_{T^2+,0}^{(j+\alpha_1,\alpha_{\tau})}(-z)
=\psi_{T^2+,0}^{(M-(j+\alpha_1),-\alpha_{\tau})}(z)
=e^{-4\pi i(j+\alpha_1)\alpha_{\tau}/M}\psi_{T^2+,0}^{(M-(j+\alpha_1),\alpha_{\tau})}(z),
\label{eq:Z2-property}
\end{align}
with $(\alpha_1,\alpha_{\tau})=(0,0),({1\over 2},0),(0,{1\over 2}),({1\over 2},{1\over 2})~(\mathrm{mod}~1)$.
Substituting this relation into eq.(\ref{T2Z2_Psi_eo}), we obtain 
\begin{align}
&\psi_{T^2/Z_2+,0}^{(j+\alpha_1,\alpha_{\tau})}(z)_{\pm 1} =
\mathcal{N}^{(j)}_{+,\pm 1}\left(\psi_{T^2+,0}^{(j+\alpha_1,\alpha_{\tau})}(z)\pm e^{-4\pi i(j+\alpha_1)\alpha_{\tau}/M}
\psi_{T^2+,0}^{(M-(j+\alpha_1),\alpha_{\tau})}(z)\right).
\end{align}

For example, for $M=\mathrm{even}$ and
$(\alpha_1,\alpha_{\tau})=(0,0)$, $\psi_{T^2/Z_2+,0}^{(j, 0)}(z)_{+1}$
satisfy the relations 
$\psi_{T^2/Z_2+,0}^{(j, 0)}(z)_{+1}=\psi_{T^2/Z_2+,0}^{(M-j, 0)}(z)_{+1}$ for $j=0,1,\cdots ,M-1$.
This implies that the zero-mode eigenstates $\psi_{T^2/Z_2+,0}^{(j,
  0)}(z)_{+1}$ only for $j=0,1,2,\cdots,{M\over 2}$ are linearly independent, and their number is equal to ${M \over 2}+1$.
On the other hand, for $M=\mathrm{even}$ and
$(\alpha_1,\alpha_{\tau})=(0,0)$, $\psi_{T^2/Z_2+,0}^{(j, 0)}(z)_{-1}$
satisfy 
the relations $\psi_{T^2/Z_2+,0}^{(j, 0)}(z)_{-1}=-\psi_{T^2/Z_2+,0}^{(M-j, 0)}(z)_{-1}$ for $j=0,1,\cdots ,M-1$.
Then, the zero-mode eigenstates $\psi_{T^2/Z_2+,0}^{(j, 0)}(z)_{-1}$
only for $j=1,2,\cdots,{M\over 2}-1$ are linearly independent, and their number is equal to ${M \over 2}-1$.

In the same way, we can obtain the number of linearly independent
zero-mode eigenstates for any $M$ and
$(\alpha_1,\alpha_{\tau})$. Results are shown in table~\ref{Table_T2Z2Psi}.
We should understand that
$\psi_{T^2/Z_2+,0}^{(j+\alpha_1,\alpha_{\tau})}(z)_{\pm 1}$
$(\psi_{T^2/Z_2-,0}^{(j+\alpha_1,\alpha_{\tau})}(z)_{\pm 1})$ exist 
only for $M>0~(M<0)$ in table~\ref{Table_T2Z2Psi}.
The number of linearly independent zero-mode eigenstates depend on
the evenness or oddness
of $M$ as well as the Scherk-Schwarz phases
$(\alpha_1,\alpha_{\tau})$, and that the sum of the numbers of
$\psi_{T^2/Z_2\pm,0}^{(j+\alpha_1,\alpha_{\tau})}(z)_{+1}$ and
$\psi_{T^2/Z_2\pm,0}^{(j+\alpha_1,\alpha_{\tau})}(z)_{-1}$ is equal to
$|M|$, 
which is the number of zero-mode wave functions $\psi_{T^2\pm,0}^{(j+\alpha_1,\alpha_{\tau})}(z)$.
It is important to note that candidates of three-generation models
can be obtained by taking $|M|=4,5,6,7,8$ with appropriate
Scherk-Schwarz phases 
in table~\ref{Table_T2Z2Psi}.

\begin{table}[htbp]
\vspace{7mm}
\begin{center}
\begin{tabular}{|c|c|c|c|}
\hline
$(\alpha_1,\alpha_{\tau})$ & $M$ & $\psi_{T^2/Z_2\pm,0}^{(j+\alpha_1,\alpha_{\tau})}(z)_{+1}$ & $\psi_{T^2/Z_2\pm,0}^{(j+\alpha_1,\alpha_{\tau})}(z)_{-1}$ \\
\hline\hline
\multirow{4}{*}{$(0,0)$} & \multirow{2}{*}{even} & \multirow{2}{*}{${|M| \over 2}+1$} & \multirow{2}{*}{${|M| \over 2}-1$}  \\
&&&\\
\cline{2-4}
& \multirow{2}{*}{odd} & \multirow{2}{*}{${|M|+1 \over 2}$} & \multirow{2}{*}{${|M|-1 \over 2}$}  \\
&&&\\
\hline
\multirow{4}{*}{$({1\over 2},0)$} & \multirow{2}{*}{even} & \multirow{2}{*}{${|M| \over 2}$} & \multirow{2}{*}{${|M| \over 2}$}  \\
&&&\\
\cline{2-4}
& \multirow{2}{*}{odd} & \multirow{2}{*}{${|M|+1 \over 2}$} & \multirow{2}{*}{${|M|-1 \over 2}$}  \\
&&&\\
\hline
\multirow{4}{*}{$(0,{1\over 2})$} & \multirow{2}{*}{even} & \multirow{2}{*}{${|M| \over 2}$} & \multirow{2}{*}{${|M| \over 2}$}  \\
&&&\\
\cline{2-4}
& \multirow{2}{*}{odd} & \multirow{2}{*}{${|M|+1 \over 2}$} & \multirow{2}{*}{${|M|-1 \over 2}$}  \\
&&&\\
\hline
\multirow{4}{*}{$({1\over 2},{1\over 2})$} & \multirow{2}{*}{even} & \multirow{2}{*}{${|M| \over 2}$} & \multirow{2}{*}{${|M| \over 2}$}  \\
&&&\\
\cline{2-4}
& \multirow{2}{*}{odd} & \multirow{2}{*}{${|M|-1 \over 2}$} & \multirow{2}{*}{${|M|+1 \over 2}$}  \\
&&&\\
\hline
\end{tabular}
\end{center}
\vspace{3mm}
\caption{The number of linearly independent zero-mode eigenstates for each $Z_{2}$ eigenvalue.}
\label{Table_T2Z2Psi}
\end{table}

\subsection{$T^2/Z_3$, $T^2/Z_4$ and $T^2/Z_6$}

In section~\ref{zero mode on T2/Z2}, we have succeeded in obtaining the linearly independent $Z_2$ eigenstates on $T^2/Z_2$.
In this section, we extend such analysis into $T^2/Z_3$, $T^2/Z_4$ and $T^2/Z_6$.

As discussed in section~\ref{sec:Z_Neigenstate}, the zero-mode
eigenstates
$\psi_{T^2/Z_N+,0}^{(j+\alpha_1,\alpha_{\tau})}(z)_{\omega^{\ell}}$
with the $Z_N$ eigenvalue $\omega^{\ell}~(\ell =0,1,\cdots, N-1)$ and
$M>0$ will be given, in terms of the zero-mode functions 
$\psi_{T^2+,0}^{(j+\alpha_1,\alpha_{\tau})}(z)$ on $T^2$, as
\begin{align}
\psi_{T^2/Z_N+,0}^{(j+\alpha_1,\alpha_{\tau})}(z)_{\omega^{\ell}} 
=\mathcal{N}^{(j)}_{+,\omega^{\ell}}\sum_{k =0}^{N-1}\bar{\omega}^{\ell k} \psi_{T^2+,0}^{(j+\alpha_1,\alpha_{\tau})}(\omega^{k}z),
\label{psi_T2/ZN+,0}
\end{align}
which obey the eigenvalue equations
\begin{align}
\psi_{T^2/Z_N+,0}^{(j+\alpha_1,\alpha_{\tau})}(\omega
z)_{\omega^{\ell}}
=\omega^{\ell}\psi_{T^2/Z_N+,0}^{(j+\alpha_1,\alpha_{\tau})}(z)_{\omega^{\ell}} ,
\label{eigenEq+0}
\end{align}
for $j=0,1,\cdots ,M-1$ and $\ell =0,1,\cdots , N-1$.

As pointed out in section~\ref{sec:Z_Neigenstate}, all of
$\psi_{T^2/Z_N+,0}^{(j+\alpha_1,\alpha_{\tau})}(z)_{\omega^{\ell}}$
for $j=0,1,\cdots,M-1$ with a fixed $\ell$ 
are not always linearly independent.
To find the number of linearly independent zero-mode eigenstates
$\psi_{T^2/Z_N+,0}^{(j+\alpha_1,\alpha_{\tau})}(z)_{\omega^{\ell}}$,
we need information on the relations between
$\psi_{T^2+,0}^{(j+\alpha_1,\alpha_{\tau})}(\omega^kz)$ and
$\psi_{T^2+,0}^{(j+\alpha_1,\alpha_{\tau})}(z)$ such as eq.~(\ref{eq:Z2-property}).
Since $\psi_{T^2+,0}^{(j+\alpha_1,\alpha_{\tau})}(\omega^kz)$ for any
$j$ and $k$ satisfies the same zero-mode equations and boundary
conditions on $T^2$ as $\psi_{T^2+,0}^{(i+\alpha_1,\alpha_{\tau})}(z)$
for $i=0,1,\cdots,M-1$, and since
$\{\psi_{T^2+,0}^{(i+\alpha_1,\alpha_{\tau})}(z),~i=0,1,\cdots,M-1\}$
forms a complete set of the zero-mode eigenstates on $T^2$,
$\psi_{T^2+,0}^{(j+\alpha_1,\alpha_{\tau})}(\omega^kz)$ have to be
expressed by some linear combination of
$\psi_{T^2+,0}^{(j+\alpha_1,\alpha_{\tau})}(z)$ 
such that
\begin{align}
\psi_{T^2+,0}^{(j+\alpha_1,\alpha_{\tau})}(\omega^k z)=\sum_{i=0}^{M-1}C_{k}^{ji}\psi_{T^2+,0}^{(i+\alpha_1,\alpha_{\tau})}(z),
\label{Psi_Ckji}
\end{align}
where $C_{k}^{ji}$ are complex coefficients, $j=0,1,\cdots, M-1$, and $k=0,1,\cdots,N-1$.

Inserting eq.(\ref{Psi_Ckji}) into eq.(\ref{eigenEq+0}), we obtain 
\begin{align}
\psi_{T^2/Z_N+,0}^{(j+\alpha_1,\alpha_{\tau})}(z)_{\omega^{\ell}}=\mathcal{N}^{(j)}_{+,\omega^{\ell}}\sum_{k
  =0}^{N-1}
\sum_{i=0}^{M-1}\bar{\omega}^{\ell k}C_{k}^{ji}\psi_{T^2+,0}^{(i+\alpha_1,\alpha_{\tau})}(z).
\end{align}
Thus, any $Z_N$ eigenstate $\psi_{T^2/Z_N+,0}^{(j+\alpha_1,\alpha_{\tau})}(z)_{\omega^{\ell}}$ on $T^2/Z_N$ turns out to be expanded in terms of the zero-mode functions $\psi_{T^2+,0}^{(i+\alpha_1,\alpha_{\tau})}(z)$ on $T^2$.
This result immediately tells us that all of
$\psi_{T^2/Z_N+,0}^{(j+\alpha_1,\alpha_{\tau})}(z)_{\omega^{\ell}}$
for $j=0,1,\cdots, M-1$ and $\ell =0,1,\cdots, N-1$ are not always
linearly independent because there are only  $M$ independent wave
functions $\psi_{T^2+,0}^{(i+\alpha_1,\alpha_{\tau})}(z)$ for
$i=0,1,\cdots,M-1$ on $T^2$, but a naive counting of the functions
$\psi_{T^2/Z_N+,0}^{(j+\alpha_1,\alpha_{\tau})}(z)_{\omega^{\ell}}$
for $j=0,1,\cdots,M-1$ and $\ell =0,1,\cdots,N-1$ gives $NM$, 
which is always larger than $M$ for $N>1$.
Thus. it becomes an important task to find $M$ linearly independent
functions from
$\psi_{T^2/Z_N+,0}^{(j+\alpha_1,\alpha_{\tau})}(z)_{\omega^{\ell}}$ 
for $j=0,1,\cdots,M-1$ and for $\ell =0,1,\cdots,N-1$.

We have succeeded in obtaining the coefficients $C_k^{ji}$ for the $Z_2$ orbifold with $\omega=-1$ in section~\ref{zero mode on T2/Z2}.
However, it seems mathematically non-trivial to determine the coefficients $C_k^{ji}$ analytically for $N=3,4,6$ and any $M$, in general.
Hence, we will first try to find $C_k^{ji}$ for some small $M$, analytically.

Let us first consider the case of $M=1$ and $(\alpha_1,\alpha_{\tau})=(1/2,1/2)$.
Then, we obtain a zero-mode function $\psi_{T^2+,0}^{({1\over 2},{1\over 2})}(z)$ on $T^2$.
It is not difficult to show that $\psi_{T^2+,0}^{({1\over 2},{1\over 2})}(z)$ satisfies the relation
\begin{align}
\psi_{T^2+,0}^{({1\over 2},{1\over 2})}(\omega z)=\omega
\psi_{T^2+,0}^{({1\over 2},{1\over 2})}(z) ,
\label{psiT2_ZNeigenEq}
\end{align}
for $\omega =e^{2\pi i/N}$ with $N=3,4,6$.
To derive the above relation, we may use the following formulae of the elliptic theta functions,
\begin{align}
&\vartheta \left[
\begin{array}{c}
{1\over 2}\\ {1\over 2}
\end{array}
\right] (-z,\tau)
=-\vartheta \left[
\begin{array}{c}
{1\over 2}\\ {1\over 2}
\end{array}
\right] (z,\tau), \notag \\
&\vartheta \left[
\begin{array}{c}
{1\over 2}\\ {1\over 2}
\end{array}
\right] (z,\tau +1)
=e^{i\pi/4}\vartheta \left[
\begin{array}{c}
{1\over 2}\\ {1\over 2}
\end{array}
\right] (z,\tau), \notag \\
&\vartheta \left[
\begin{array}{c}
{1\over 2}\\ {1\over 2}
\end{array}
\right] (-{z\over \tau},-{1 \over \tau})
=i(-i\tau)^{1/2}e^{i\pi z^2/\tau}\vartheta \left[
\begin{array}{c}
{1\over 2}\\ {1\over 2}
\end{array}
\right] (z,\tau),
\end{align}
with $\tau=\omega =e^{2\pi i/N}$ for $N=3,4,6$.
It follows from eqs.(\ref{psi_T2/ZN+,0}) and (\ref{psiT2_ZNeigenEq}) that we find
\begin{align}
\psi_{T^2/Z_N+,0}^{({1\over 2},{1\over 2})}(z)_{\omega^{\ell}}
=\left\{
\begin{array}{c}
\mathcal{N}_{+,\omega}^{(0)}\psi_{T^2+,0}^{({1\over 2},{1\over 2})}(z)~~~~~\mathrm{for}~~\ell =1\\
0 \hspace{31.8mm} \mathrm{for}~~\ell\ne 1
\end{array}
\right..
\end{align}
Thus, the $Z_N$ eigenstates $\psi_{T^2/Z_N+,0}^{({1\over 2},{1\over
    2})}(z)_{\omega^{\ell}}$ defined in eq.(\ref{psi_T2/ZN+,0}) are
non-vanishing only when $\ell =1$ 
for $M=1$ and $(\alpha_1,\alpha_{\tau})=(1/2,1/2)$.

A similar analysis will work for $M=1$ and $(\alpha_1,\alpha_{\tau})=(0,0)$ with $N=4$.\footnote{
Remember that for $M=\mathrm{odd}$, the Scherk-Schwarz phase $(\alpha_1,\alpha_{\tau})=(0,0)$ is permitted only for $N=2,4$ (see table~\ref{table:SS phases}).}
Then, we can show that 
\begin{align}
\psi_{T^2+,0}^{(0,0)}(\omega z)= \psi_{T^2+,0}^{(0,0)}(z)~~~~~\mathrm{for}~~\omega =i.
\end{align}
It follows that 
\begin{align}
\psi_{T^2/Z_4+,0}^{(0,0)}(z)_{\omega^{\ell}}
=\left\{
\begin{array}{l}
\mathcal{N}_{+,+1}^{(0)}\psi_{T^2+,0}^{(0,0)}(z) \hspace{10mm}\mathrm{for}~~\ell =0\\
0 \hspace{35.5mm} \mathrm{for}~~\ell =1,2,3
\end{array}
\right..
\end{align}

Let us next discuss the zero-mode eigenstates on $T^2/Z_4$ for $M=2$ and $(\alpha_1,\alpha_{\tau})=(0,0)$.
For $M=2$, there are two zero-mode solutions $\psi_{T^2+,0}^{(j,0)}(z)$ for $j=0,1$ on $T^2$.
Using the formulae of the elliptic theta functions given in appendix~\ref{Exaple of calculation}, we can show that 
\begin{align}
&\psi_{T^2+,0}^{(0,0)}(iz) ={1\over \sqrt{2}}(\psi_{T^2+,0}^{(0,0)}(z)+\psi_{T^2+,0}^{(1,0)}(z)), \notag \\
&\psi_{T^2+,0}^{(1,0)}(iz) ={1\over \sqrt{2}}(\psi_{T^2+,0}^{(0,0)}(z)-\psi_{T^2+,0}^{(1,0)}(z)).
\end{align}
It follows that we obtain
\begin{align}
&\psi_{T^2/Z_4+,0}^{(0,0)}(z)_{1} 
=\mathcal{N}^{(0)}_{+,1} \left( \psi_{T^2+,0}^{(0,0)}(z)+(\sqrt{2}-1)\psi_{T^2+,0}^{(1,0)}(z)\right), \notag \\
&\psi_{T^2/Z_4+,0}^{(1,0)}(z)_{1} 
=\mathcal{N}^{(1)}_{+,1} \left(\psi_{T^2+,0}^{(0,0)}(z)+(\sqrt{2}-1)\psi_{T^2+,0}^{(1,0)}(z)\right), \notag \\
&\psi_{T^2/Z_4+,0}^{(0,0)}(z)_{\omega}=\psi_{T^2/Z_4+,0}^{(1,0)}(z)_{\omega}=0, \notag \\
&\psi_{T^2/Z_4+,0}^{(0,0)}(z)_{\omega^2} 
=\mathcal{N}^{(0)}_{+,\omega^2} \left(\psi_{T^2+,0}^{(0,0)}(z)-(\sqrt{2}+1)\psi_{T^2+,0}^{(1,0)}(z)\right), \notag \\
&\psi_{T^2/Z_4+,0}^{(1,0)}(z)_{\omega^2} 
=\mathcal{N}^{(1)}_{+,\omega^2} \left(\psi_{T^2+,0}^{(0,0)}(z)-(\sqrt{2}+1)\psi_{T^2+,0}^{(1,0)}(z)\right), \notag \\
&\psi_{T^2/Z_4+,0}^{(0,0)}(z)_{\omega^3}=\psi_{T^2/Z_4+,0}^{(1,0)}(z)_{\omega^3}=0.
\label{example_Z4_1}
\end{align}
Thus, the number of linearly independent zero-mode $Z_4$ eigenstates
$\psi_{T^2/Z_4+,0}^{(j,0)}(z)_{\omega^{\ell}}$ $(\ell =0,1,2,3)$ turns
out to be 
$1,0,1,0$ for $\ell =0,1,2,3$, respectively.
The total number of the linearly independent zero-mode eigenstates is
equal to two, and is identical to the number of zero-mode solutions
$\psi_{T^2+,0}^{(j,0)}(z)~(j=0,1)$ on $T^2$, 
as expected.

We present one more example of $M=4$ and
$(\alpha_1,\alpha_{\tau})=(0,0)$ on $T^2/Z_4$ to show the
non-triviality of getting $Z_N$ eigenstates on $Z_N$ orbifolds 
with magnetic flux.
The results are given as
\begin{align}
\psi_{T^2/Z_4+,0}^{(0,0)}(z)_{1}
&=\mathcal{N}^{(0)}_{+,1} \left(3\psi_{T^2+,0}^{(0,0)}(z)+\psi_{T^2+,0}^{(1,0)}(z)+\psi_{T^2+,0}^{(2,0)}(z)+\psi_{T^2+,0}^{(3,0)}(z)\right), \notag \\
\psi_{T^2/Z_4+,0}^{(1,0)}(z)_{1}
&=\psi_{T^2/Z_4+,0}^{(3,0)}(z)_{1} \notag \\
&=\mathcal{N}^{(1)}_{+,1} \left(\psi_{T^2+,0}^{(0,0)}(z)+\psi_{T^2+,0}^{(1,0)}(z)-\psi_{T^2+,0}^{(2,0)}(z)+\psi_{T^2+,0}^{(3,0)}(z)\right), \notag \\
\psi_{T^2/Z_4+,0}^{(2,0)}(z)_{1}
&=\mathcal{N}^{(2)}_{+,1} \left(\psi_{T^2+,0}^{(0,0)}(z)-\psi_{T^2+,0}^{(1,0)}(z)+3\psi_{T^2+,0}^{(2,0)}(z)-\psi_{T^2+,0}^{(3,0)}(z)\right) \notag \\
&=\mathcal{N}^{(2)}_{+,1} \left({1\over \mathcal{N}^{(0)}_{+,1}}\psi_{T^2/Z_4+,0}^{(0,0)}(z)_{1}  -{2\over \mathcal{N}^{(1)}_{+,1}}\psi_{T^2/Z_4+,0}^{(1,0)}(z)_{1}\right) \notag \\
\psi_{T^2/Z_4+,0}^{(0,0)}(z)_{\omega}
&=\psi_{T^2/Z_4+,0}^{(2,0)}(z)_{\omega}=0 \notag \\
\psi_{T^2/Z_4+,0}^{(1,0)}(z)_{\omega}
&=-\psi_{T^2/Z_4+,0}^{(3,0)}(z)_{\omega}=\mathcal{N}^{(1)}_{+,\omega} \left(\psi_{T^2+,0}^{(1,0)}(z)-\psi_{T^2+,0}^{(3,0)}(z)\right) \notag \\
\psi_{T^2/Z_4+,0}^{(0,0)}(z)_{\omega^2}
&=-\psi_{T^2/Z_4+,0}^{(1,0)}(z)_{\omega^2}=-\psi_{T^2/Z_4+,0}^{(2,0)}(z)_{\omega^2}=\psi_{T^2/Z_4+,0}^{(3,0)}(z)_{\omega^2} \notag \\
&=\mathcal{N}^{(0)}_{+,\omega^2}\left( \psi_{T^2+,0}^{(0,0)}(z)-\psi_{T^2+,0}^{(1,0)}(z)-\psi_{T^2+,0}^{(2,0)}(z)-\psi_{T^2+,0}^{(3,0)}(z)\right) , \notag \\
\psi_{T^2/Z_4+,0}^{(0,0)}(z)_{\omega^3}
&=\psi_{T^2/Z_4+,0}^{(1,0)}(z)_{\omega^3}=\psi_{T^2/Z_4+,0}^{(2,0)}(z)_{\omega^3}=\psi_{T^2/Z_4+,0}^{(3,0)}(z)_{\omega^3}=0.
\end{align}

In the above analysis, we have explicitly found the coefficients
$C_k^{ji}$ in eq.(\ref{Psi_Ckji}) for some small $M$ and 
the number of linearly independent $Z_N$ eigenstates on $T^2/Z_N~(N=3,4,6)$.
However, it is difficult to continue this analysis for larger $M$ because necessary formulae of the elliptic theta functions to determine $C_k^{ji}$ are not known to the best of our knowledge.

Therefore, we will determine the coefficients $C_k^{ji}$ and the
numbers of linearly independent $Z_N$ eigenstates
$\psi_{T^2/Z_N\pm,0}^{(j+\alpha_1,\alpha_{\tau})}(z)_{\omega^{\ell}}$
on $T^2/Z_N~(N=3,4,6)$ for larger $M$ by numerical 
analysis.\footnote{
The numerical values of $C_k^{ji}$ will be confirmed by another approach of the operator formalism given in ref.\cite{Abe:2013}.} 
The results for $T^2/Z_3$, $T^2/Z_4$ and $T^2/Z_6$ are given in tables~\ref{T2Z3Psi_1e}-\ref{T2Z6Psi_2}.
Those tables show the number of linearly independent $Z_N$
eigenfunctions
$\psi_{T^2/Z_N\pm,0}^{(j+\alpha_1,\alpha_{\tau})}(z)_{\eta}$ for each
combination of $\eta=\omega^{\ell}~(\ell =0,1,\cdots,N-1)$ 
and $|M|$.
For example, when we want to construct a three-generation model on
$T^2/Z_3$, we may choose one of
$(|M|,\eta)=(6,1),(8,1),(10,1),(8,\bar{\omega}),(10,\bar{\omega}),(12,\bar{\omega})$ 
with $\omega =e^{2\pi i/3}$ in table~\ref{T2Z3Psi_1e}.

\begin{figure}[htbp]
\vspace{20mm}
\begin{minipage}{0.5\hsize}
\begin{center}
\includegraphics[width=70mm]{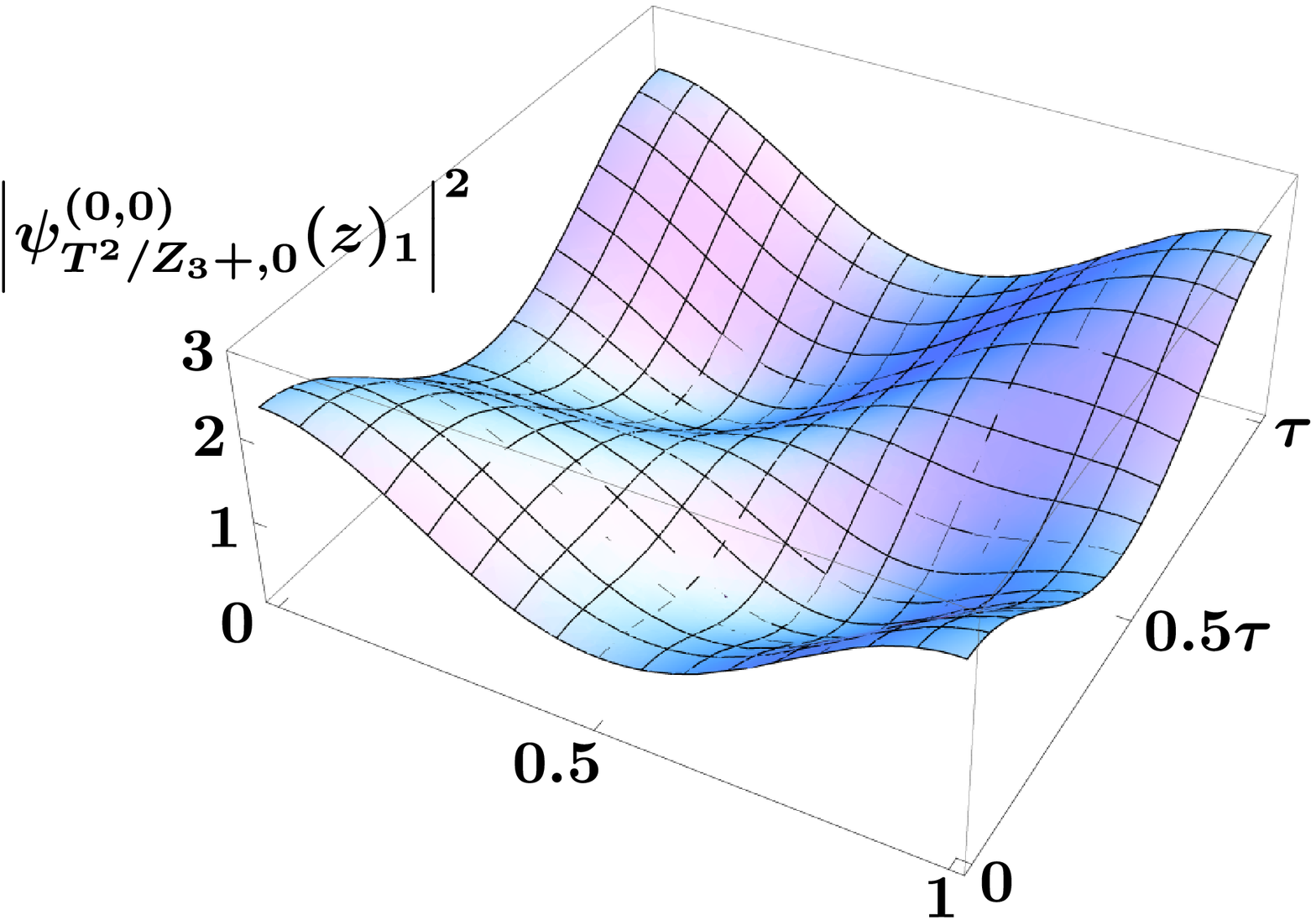}
\includegraphics[width=70mm]{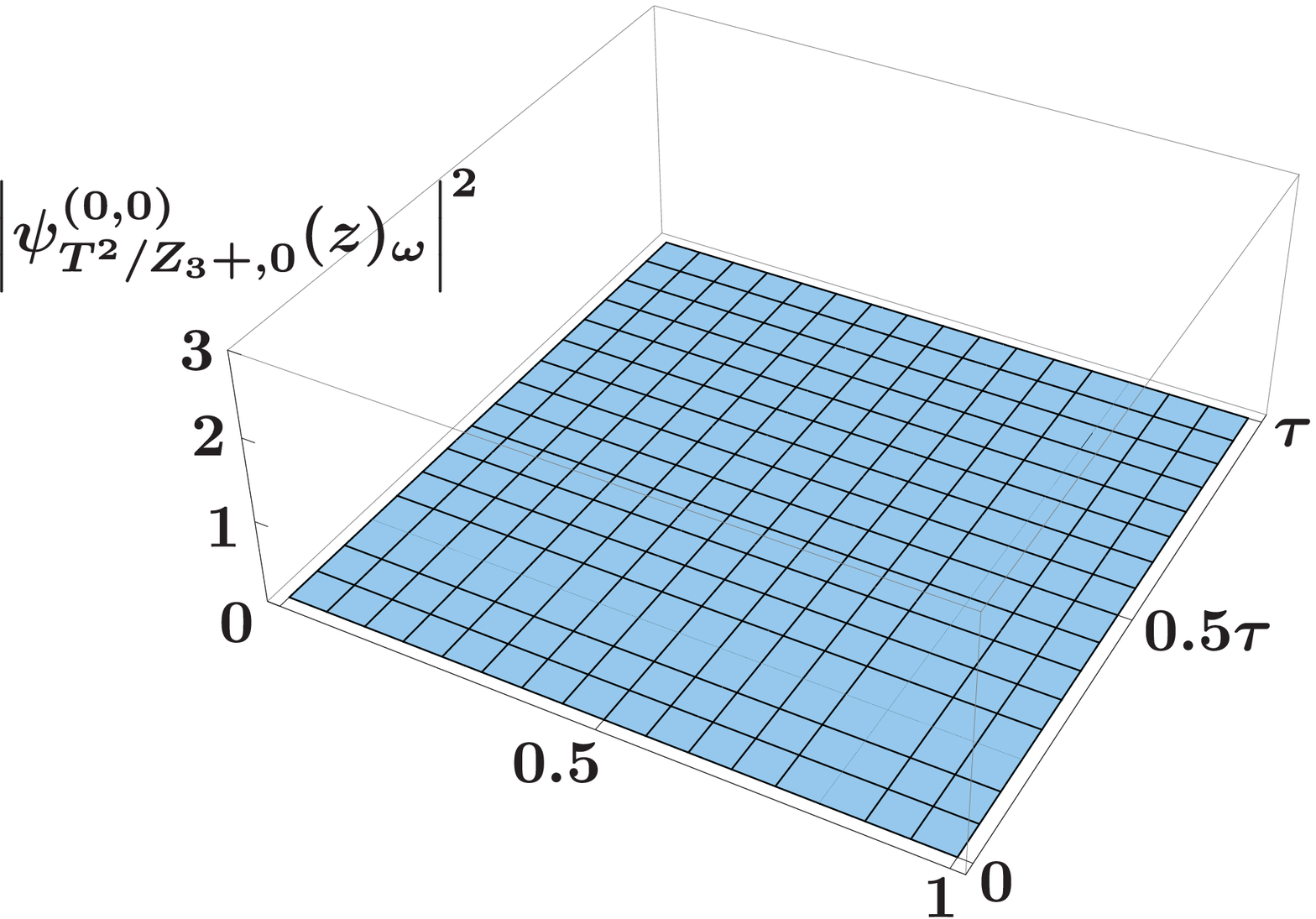}
\includegraphics[width=70mm]{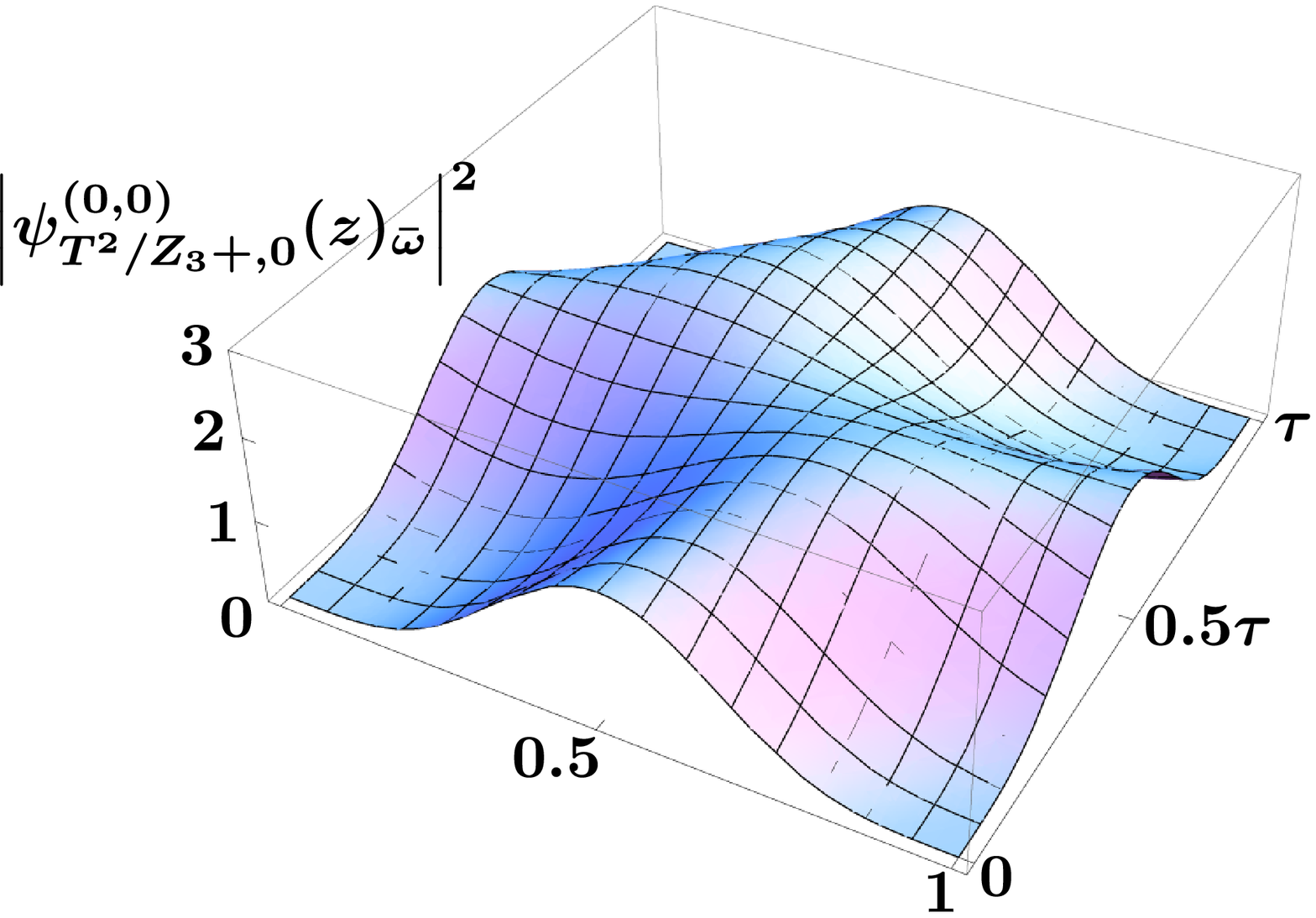}
\end{center}
\end{minipage}
\begin{minipage}{0.49\hsize}
\begin{center}
\includegraphics[width=70mm]{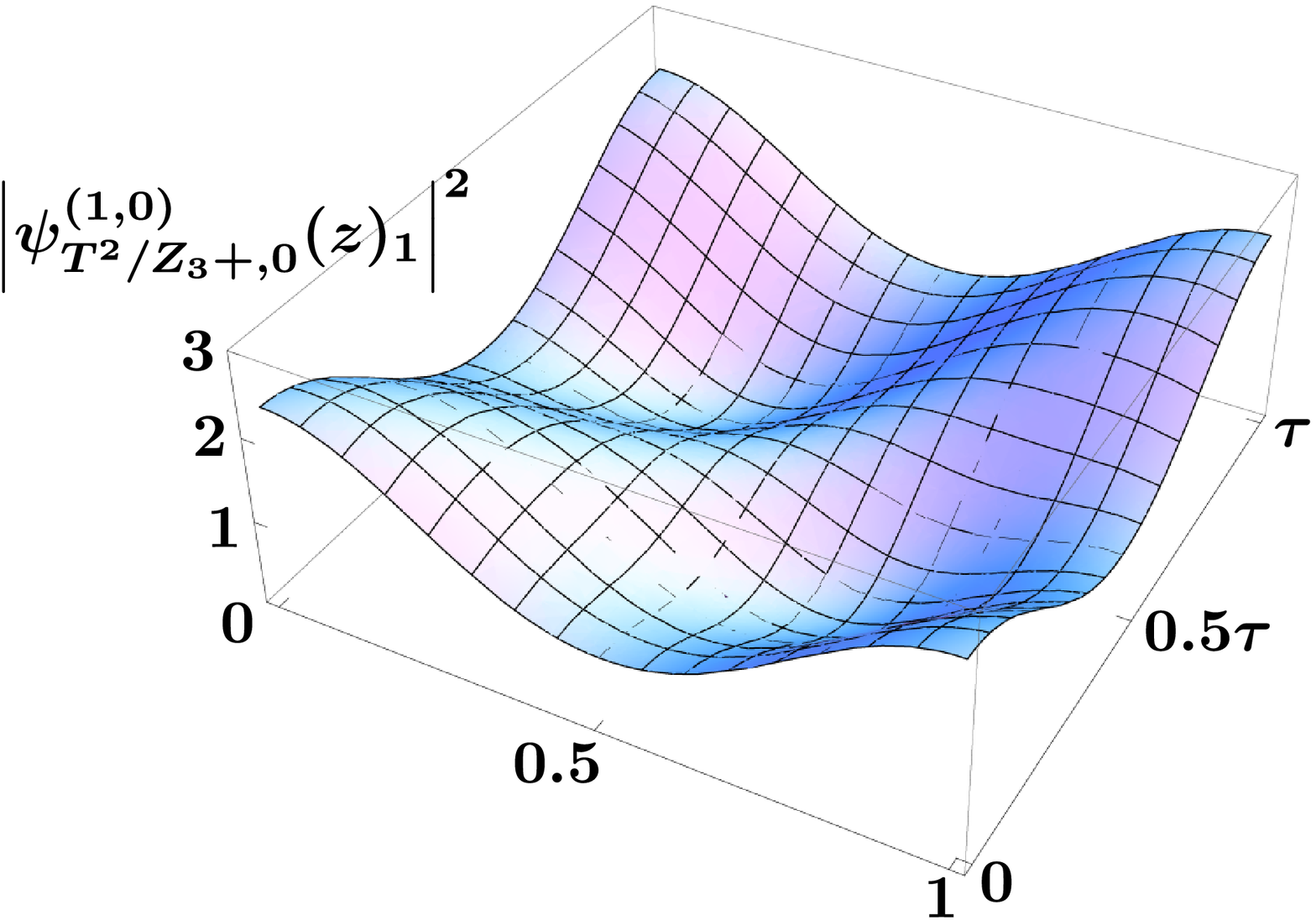}
\includegraphics[width=70mm]{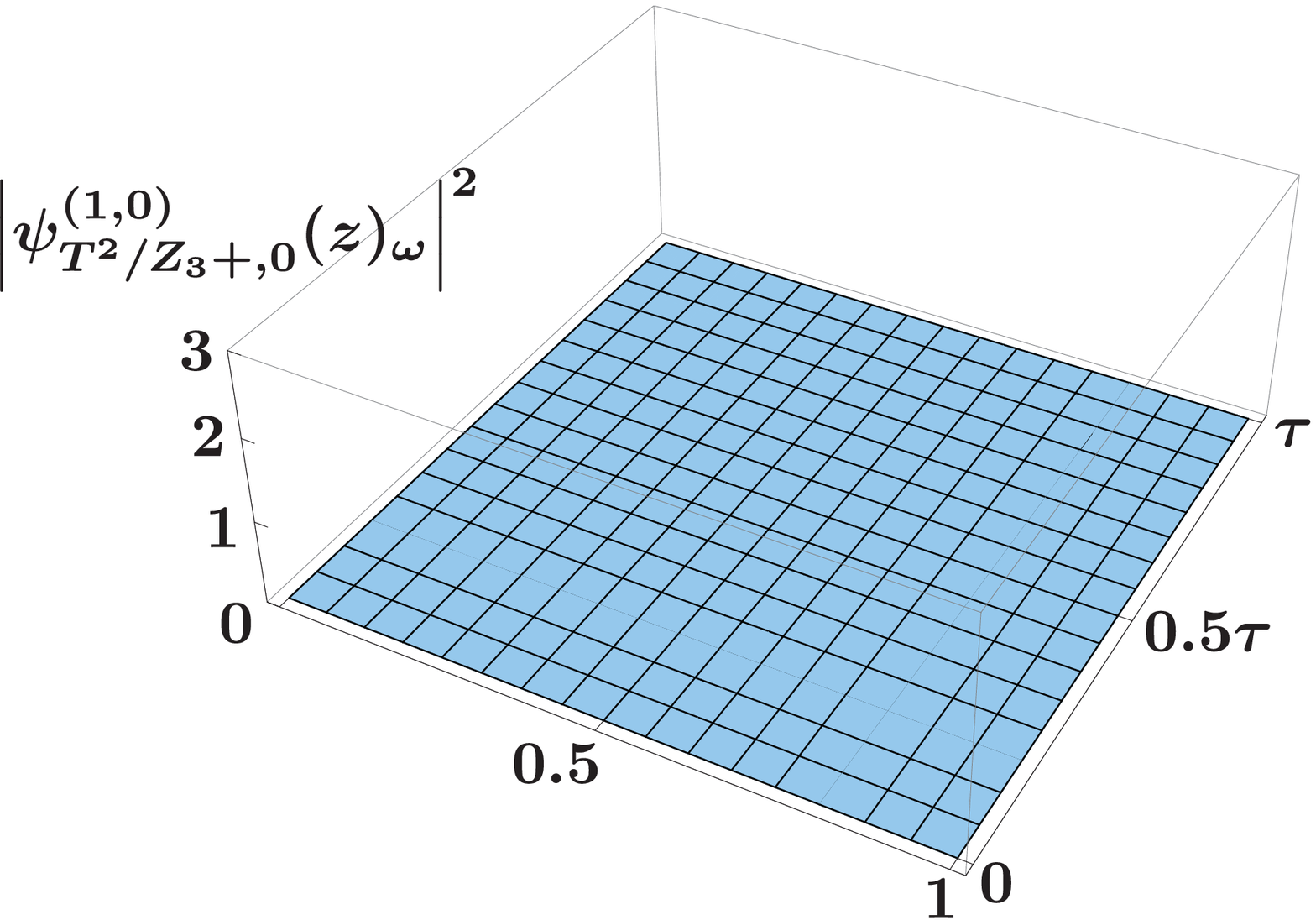}
\includegraphics[width=70mm]{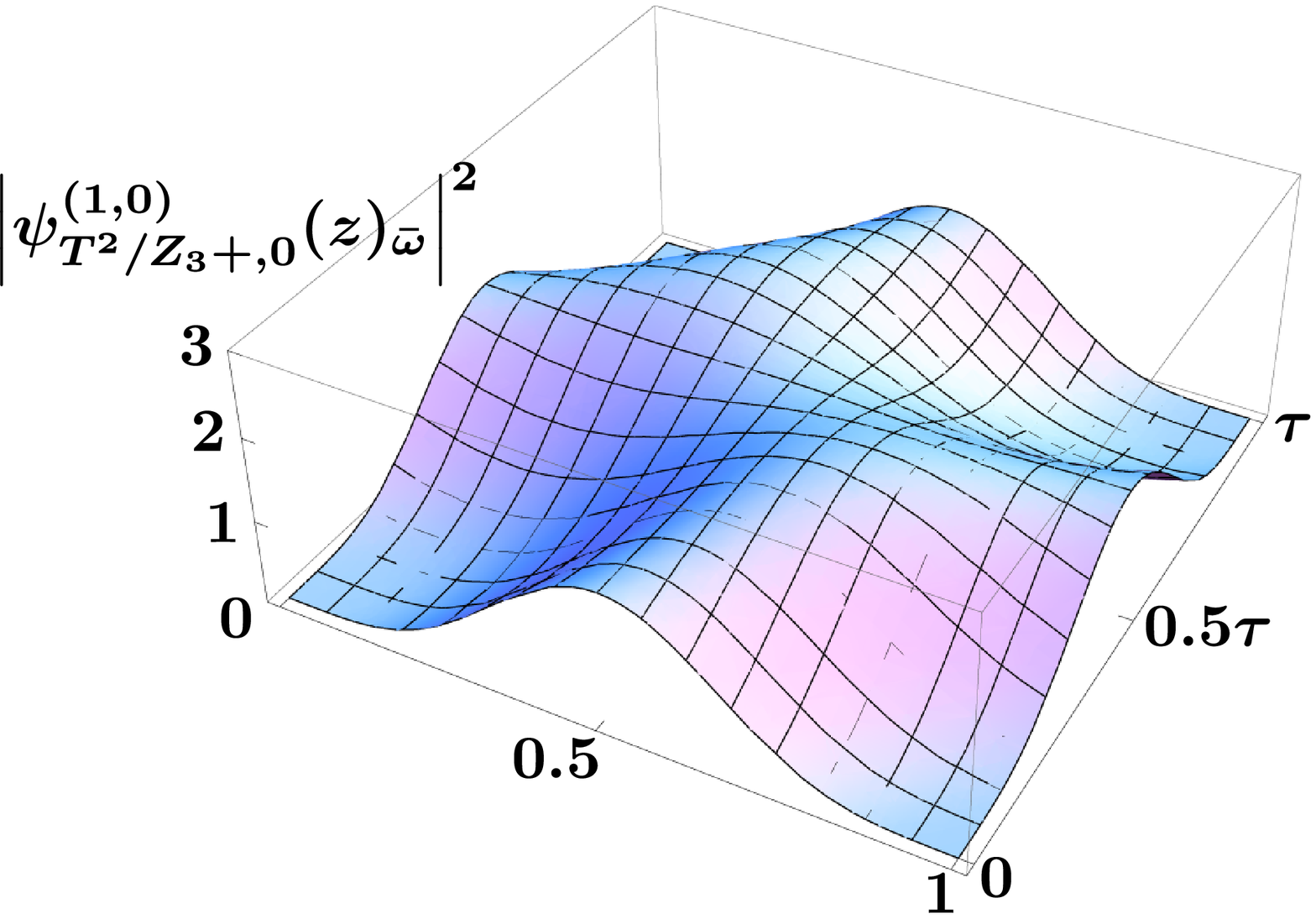}
\end{center}
\end{minipage}
\caption{A numerical analysis of the probability densities $|\psi_{T^2/Z_3+,0}^{(j,0)}(z)_{\omega^k}|^2$ for $M=2$.
These figures are depicted with a complex coordinate $z=x+\tau y~(0\leq x,y \leq 1)$ for convenience of explanation.
The top two figures are probability densities
$|\psi_{T^2/Z_3+,0}^{(j,0)}(z)_{1}|^2$, the middle two figures are
$|\psi_{T^2/Z_3+,0}^{(j,0)}(z)_{\omega}|^2$, 
and the bottom two figures are $|\psi_{T^2/Z_3+,0}^{(j,0)}(z)_{\bar{\omega}}|^2$. 
The left side is $|\psi_{T^2/Z_3+,0}^{(0,0)}(z)_{\eta}|^2$, and the right side is $|\psi_{T^2/Z_3+,0}^{(1,0)}(z)_{\eta}|^2$.}
\label{fig:direct Z3 M=2}
\vspace{14mm}
\end{figure}

As an example of numerical computations, numerical values of
$|\psi_{T^2/Z_3+,0}^{(j,0)}(z)_{\omega^{\ell}}|^2$ for $M=2$ and 
$(\alpha_1,\alpha_{\tau})=(0,0)$ are depicted in figure~\ref{fig:direct Z3 M=2}.
The two figures of the top, middle and bottom in figure~\ref{fig:direct Z3
  M=2} correspond to $|\psi_{T^2/Z_3+,0}^{(j,0)}(z)_{1}|^2$,
$|\psi_{T^2/Z_3+,0}^{(j,0)}(z)_{\omega}|^2$ 
and $|\psi_{T^2/Z_3+,0}^{(j,0)}(z)_{\bar{\omega}}|^2$, respectively.
The figures of the left (right) side are $|\psi_{T^2/Z_3+,0}^{(0,0)}(z)_{\omega^{\ell}}|^2$ $(|\psi_{T^2/Z_3+,0}^{(1,0)}(z)_{\omega^{\ell}}|^2)$.
The wave functions $\psi_{T^2/Z_3+,0}^{(j,0)}(z)_{\omega^{\ell}}$ have been obtained by computing the right-hand side of eq.(\ref{psi_T2/ZN+,0}) numerically.
We have also determined the coefficients $C_k^{ji}$ in
eq.(\ref{Psi_Ckji}) numerically, and confirmed that
eq.(\ref{psi_T2/ZN+,0}) 
is consistent with eq.(\ref{Psi_Ckji}).
It follows from figure~\ref{fig:direct Z3 M=2} that
$\psi_{T^2/Z_3+,0}^{(1,0)}(z)_{1}~(\psi_{T^2/Z_3+,0}^{(1,0)}(z)_{\bar{\omega}})$
is linearly dependent on 
$\psi_{T^2/Z_3+,0}^{(0,0)}(z)_{1}~(\psi_{T^2/Z_3+,0}^{(0,0)}(z)_{\bar{\omega}})$
, while $\psi_{T^2/Z_3+,0}^{(0,0)}(z)_{\omega}=\psi_{T^2/Z_3+,0}^{(1,0)}(z)_{\omega}=0$ in the numerical analysis.
The numerical results agree with those derived analytically before.
\begin{table}[htbp]
\vspace{8mm}
\begin{center}
\begin{tabular}{|c|c|ccccccc|}
\hline
\multicolumn{2}{|c|}{$|M|$}& 2 & 4 & 6 & 8 & 10 & 12 & 14 \\
\hline
\multirow{3}{*}{$\eta$}&1 & 1 & 1 & 3 & 3 & 3 & 5 & 5 \\
&$\omega$ & 0 & 2 & 2 & 2 & 4 & 4 & 4 \\
&$\bar{\omega}$ & 1 & 1 & 1 & 3 & 3 & 3 & 5 \\
\hline
\end{tabular}
\end{center}
\vspace{-3mm}
\caption{The number of linearly independent zero-mode eigenstates
  $\psi_{T^2/Z_3\pm,0}(z)_{\eta}$ for $M=\mathrm{even}$ and
  $(\alpha_1,\alpha_{\tau})=(0, 0)$ 
on $T^2/Z_3$.}
\label{T2Z3Psi_1e}
\end{table}
\begin{table}[htbp]
\vspace{12mm}
\begin{center}
\begin{tabular}{|c|c|ccccccc|}
\hline
\multicolumn{2}{|c|}{$|M|$}& 2 & 4 & 6 & 8 & 10 & 12 & 14 \\
\hline
\multirow{3}{*}{$\eta$}&1 & 1 & 2 & 2 & 3 & 4 & 4 & 5 \\
&$\omega$ & 1 & 1 & 2 & 3 & 3 & 4 & 5 \\
&$\bar{\omega}$ & 0 & 1 & 2 & 2 & 3 & 4 & 4 \\
\hline
\end{tabular}
\end{center}
\vspace{-3mm}
\caption{The number of linearly independent zero-mode eigenstates
  $\psi_{T^2/Z_3\pm,0}(z)_{\eta}$ for $M=\mathrm{even}$ and
  $(\alpha_1,\alpha_{\tau})=\left({1\over 3},{1\over
      3}\right),\left({2\over 3},{2\over 3}\right)$ 
on $T^2/Z_3$.}
\label{T2Z3Psi_2e}
\end{table}
\begin{table}[htbp]
\vspace{12mm}
\begin{center}
\begin{tabular}{|c|c|ccccccc|}
\hline
\multicolumn{2}{|c|}{$|M|$}& 1 & 3 & 5 & 7 & 9 & 11 & 13 \\
\hline
\multirow{3}{*}{$\eta$}&1 & 1 & 1 & 2 & 3 & 3 & 4 & 5 \\
&$\omega$ & 0 & 1 & 2 & 2 & 3 & 4 & 4 \\
&$\bar{\omega}$ & 0 & 1 & 1 & 2 & 3 & 3 & 4 \\
\hline
\end{tabular}
\end{center}
\vspace{-3mm}
\caption{The number of linearly independent zero-mode eigenstates
  $\psi_{T^2/Z_3\pm,0}(z)_{\eta}$ for $M=\mathrm{odd}$ and
  $(\alpha_1,\alpha_{\tau})=\left({1\over 6},{1\over
      6}\right),\left({5\over 6},{5\over 6}\right)$ 
on $T^2/Z_3$.}
\label{T2Z3Psi_1o}
\end{table}
\begin{table}[!htbp]
\vspace{12mm}
\begin{center}
\begin{tabular}{|c|c|ccccccc|}
\hline
\multicolumn{2}{|c|}{$|M|$}& 1 & 3 & 5 & 7 & 9 & 11 & 13 \\
\hline
\multirow{3}{*}{$\eta$}&1 & 0 & 2 & 2 & 2 & 4 & 4 & 4 \\
&$\omega$ & 1 & 1 & 1 & 3 & 3 & 3 & 5 \\
&$\bar{\omega}$ & 0 & 0 & 2 & 2 & 2 & 4 & 4 \\
\hline
\end{tabular}
\end{center}
\vspace{-3mm}
\caption{The number of linearly independent zero-mode eigenstates
  $\psi_{T^2/Z_3\pm,0}(z)_{\eta}$ for $M=\mathrm{odd}$ and
  $(\alpha_1,\alpha_{\tau})=\left({1\over 2},{1\over 2}\right)$ 
on $T^2/Z_3$.}
\label{T2Z3Psi_2o}
\end{table}


\begin{table}[htbp]
\vspace{4mm}
\begin{center}
\begin{tabular}{|c|c|ccccccccccccccccc|}
\hline
\multicolumn{2}{|c|}{$|M|$}& 1 & 2 & 3 & 4 & 5 & 6 & 7 & 8 & 9 & 10 & 11 & 12 & 13 & 14 & 15 & 16 & 17  \\
\hline
\multirow{4}{*}{$\eta$}&$+1$ & 1 & 1 & 1 & 2 & 2 & 2 & 2 & 3 & 3 & 3 & 3 & 4 & 4 & 4 & 4 & 5 & 5 \\
&$+i$ & 0 & 0 & 1 & 1 & 1 & 1 & 2 & 2 & 2 & 2 & 3 & 3 & 3 & 3 & 4 & 4 & 4 \\
&$-1$ & 0 & 1 & 1 & 1 & 1 & 2 & 2 & 2 & 2 & 3 & 3 & 3 & 3 & 4 & 4 & 4 & 4 \\
&$-i$ & 0 & 0 & 0 & 0 & 1 & 1 & 1 & 1 & 2 & 2 & 2 & 2 & 3 & 3 & 3 & 3 & 4 \\
\hline
\end{tabular}
\end{center}
\vspace{-3mm}
\caption{The number of linearly independent zero-mode eigenstates $\psi_{T^2/Z_4\pm,0}(z)_{\eta}$ for $(\alpha_1,\alpha_{\tau})=(0, 0)$ on $T^2/Z_4$.}
\label{T2Z4Psi_1}
\end{table}
\begin{table}[htbp]
\vspace{6mm}
\begin{center}
\begin{tabular}{|c|c|cccccccccccccccc|}
\hline
\multicolumn{2}{|c|}{$|M|$}& 1 & 2 & 3 & 4 & 5 & 6 & 7 & 8 & 9 & 10 & 11 & 12 & 13 & 14 & 15 & 16  \\
\hline
\multirow{4}{*}{$\eta$}&$+1$ & 0 & 1 & 1 & 1 & 1 & 2 & 2 & 2 & 2 & 3 & 3 & 3 & 3 & 4 & 4 & 4 \\
&$+i$ & 1 & 1 & 1 & 1 & 2 & 2 & 2 & 2 & 3 & 3 & 3 & 3 & 4 & 4 & 4 & 4 \\
&$-1$ & 0 & 0 & 0 & 1 & 1 & 1 & 1 & 2 & 2 & 2 & 2 & 3 & 3 & 3 & 3 & 4 \\
&$-i$ & 0 & 0 & 1 & 1 & 1 & 1 & 2 & 2 & 2 & 2 & 3 & 3 & 3 & 3 & 4 & 4 \\
\hline
\end{tabular}
\end{center}
\vspace{-3mm}
\caption{The number of linearly independent zero-mode eigenstates
  $\psi_{T^2/Z_4\pm,0}(z)_{\eta}$ for
  $(\alpha_1,\alpha_{\tau})=\left({1\over 2},{1\over 2}\right)$ 
on $T^2/Z_4$.}
\label{T2Z4Psi_2}
\end{table}
\begin{table}[htbp]
\vspace{6mm}
\begin{center}
\begin{tabular}{|c|c|ccccccccccccc|}
\hline
\multicolumn{2}{|c|}{$|M|$}& 2 & 4 & 6 & 8 & 10 & 12 & 14 & 16 & 18 & 20 & 22 & 24 & 26 \\
\hline
\multirow{6}{*}{$\eta$}&1 & 1 & 1 & 2 & 2 & 2 & 3 & 3 & 3 & 4 & 4 & 4 & 5 & 5 \\
&$\omega$ & 0 & 1 & 1 & 1 & 2 & 2 & 2 & 3 & 3 & 3 & 4 & 4 & 4 \\
&$\omega^2$ & 1 & 1 & 1 & 2 & 2 & 2 & 3 & 3 & 3 & 4 & 4 & 4 & 5 \\
&$\omega^3$ & 0 & 0 & 1 & 1 & 1 & 2 & 2 & 2 & 3 & 3 & 3 & 4 & 4 \\
&$\omega^4$ & 0 & 1 & 1 & 1 & 2 & 2 & 2 & 3 & 3 & 3 & 4 & 4 & 4 \\
&$\omega^5$ & 0 & 0 & 0 & 1 & 1 & 1 & 2 & 2 & 2 & 3 & 3 & 3 & 4 \\
\hline
\end{tabular}
\end{center}
\vspace{-3mm}
\caption{The number of linearly independent zero-mode eigenstates
  $\psi_{T^2/Z_6\pm,0}(z)_{\eta}$ for $M=\mathrm{even}$ and
  $(\alpha_1,\alpha_{\tau})=(0, 0)$ 
on $T^2/Z_6$.}
\label{T2Z6Psi_1}
\end{table}
\begin{table}[htbp]
\vspace{6mm}
\begin{center}
\begin{tabular}{|c|c|ccccccccccccc|}
\hline
\multicolumn{2}{|c|}{$|M|$}& 1 & 3 & 5 & 7 & 9 & 11 & 13 & 15 & 17 & 19 & 21 & 23 & 25 \\
\hline
\multirow{6}{*}{$\eta$}&1 & 0 & 1 & 1 & 1 & 2 & 2 & 2 & 3 & 3 & 3 & 4 & 4 & 4 \\
&$\omega$ & 1 & 1 & 1 & 2 & 2 & 2 & 3 & 3 & 3 & 4 & 4 & 4 & 5 \\
&$\omega^2$ & 0 & 0 & 1 & 1 & 1 & 2 & 2 & 2 & 3 & 3 & 3 & 4 & 4 \\
&$\omega^3$ & 0 & 1 & 1 & 1 & 2 & 2 & 2 & 3 & 3 & 3 & 4 & 4 & 4 \\
&$\omega^4$ & 0 & 0 & 0 & 1 & 1 & 1 & 2 & 2 & 2 & 3 & 3 & 3 & 4 \\
&$\omega^5$ & 0 & 0 & 1 & 1 & 1 & 2 & 2 & 2 & 3 & 3 & 3 & 4 & 4 \\
\hline
\end{tabular}
\end{center}
\vspace{-3mm}
\caption{The number of linearly independent zero-mode eigenstates
  $\psi_{T^2/Z_6\pm,0}(z)_{\eta}$ for $M=\mathrm{odd}$ and
  $(\alpha_1,\alpha_{\tau})=\left({1\over 2},{1\over 2}\right)$ 
on $T^2/Z_6$.}
\label{T2Z6Psi_2}
\end{table}

\newpage

\section{Kaluza-Klein mode functions and mass spectra \label{sec:KK and masses}}

In the previous section, we have considered zero-mode solutions on $T^2/Z_N$.
It is also worthwhile to discuss Kaluza-Klein modes on $T^2$ and $T^2/Z_N$.\footnote{
For a detailed investigation of the gauge theory on the torus, see ref.\cite{Cremades:2004wa,BerasaluceGonzalez:2012vb,Hamada:2012wj}}
Then, we can understand the Kaluza-Klein modes by a way similar to the analysis of the harmonic oscillator in the quantum mechanics, as we will see below.   

{}From eqs.(\ref{6D_Dirac to two Weyl}) and (\ref{mass_equation}), the masses squared of Kaluza-Klein modes with $a_w=0$ on $M^4\times T^2$ are given by
\begin{align}
\left(
\begin{array}{cc}
-4D_z^{(b)}D_{\bar{z}}^{(b)} & 0 \\
0 & -4D_{\bar{z}}^{(b)}D_{z}^{(b)}
\end{array}
\right)\left(
\begin{array}{c}
\psi_{T^2+,n}^{(j+\alpha_1,\alpha_{\tau})}(z) \\ \psi_{T^2-,n}^{(j+\alpha_1,\alpha_{\tau})}(z)
\end{array}
\right) =m_n^2\left(
\begin{array}{cc}
\psi_{T^2+,n}^{(j+\alpha_1,\alpha_{\tau})}(z) \\ \psi_{T^2-,n}^{(j+\alpha_1,\alpha_{\tau})}(z)
\end{array}
\right),
\end{align}
where $D_z^{(b)}=\partial_z-iqA_z^{(b)}(z)$ and $D_{\bar{z}}^{(b)}=\partial_{\bar{z}}-iqA_{\bar{z}}^{(b)}(z)$.
Here, we define the two-dimensional Laplace operator $\Delta$ as  
\begin{align}
\Delta \equiv -2(D_z^{(b)}D_{\bar{z}}^{(b)}+D_{\bar{z}}^{(b)}D_z^{(b)}),
\end{align}
which satisfies the relations with $D_z^{(b)}$ and $D_{\bar{z}}^{(b)}$
\begin{align}
[\Delta ,D_z^{(b)}]={4\pi M \over \mathcal{A}}D_z^{(b)},~~~[\Delta
,D_{\bar{z}}^{(b)}]=-{4\pi M \over
  \mathcal{A}}D_{\bar{z}}^{(b)},~~~[D_z^{(b)},D_{\bar{z}}^{(b)}]
={\pi M\over \mathcal{A}},
\end{align}
where $\mathcal{A}~(=\mathrm{Im}\tau\cdot 1)$ is the area of the torus.
This algebra of operators 
for $\psi_{T^2\pm,n}(z)$ is similar to 
the one of the one-dimensional harmonic oscillator in quantum mechanics.
Indeed, it is found that 
for $M>0$, 
\begin{align}
&\Delta ={4\pi |M| \over \mathcal{A}}\left(\hat{N}_+ +{1\over 2} \right), ~~~\hat{N}_+\equiv \hat{a}_+^{\dag}\hat{a}_+, \notag \\ 
&\hat{a}_+\equiv i\sqrt{\mathcal{A} \over \pi
  |M|}D_{\bar{z}}^{(b)},~~~\hat{a}_+^{\dag}\equiv i\sqrt{\mathcal{A}
  \over \pi |M|}
D_{z}^{(b)},~~~[\hat{a}_+,\hat{a}_+^{\dag}]=1, 
\label{1DHOinQM_+}
\end{align}
with
\begin{align}
\Bigl|0,{j+\alpha_1 \over M},\alpha_{\tau}\Bigr\rangle_{T^2+} \equiv \psi_{T^2+,0}^{(j+\alpha_1,\alpha_{\tau})}(z), 
\end{align}
and for $M<0$, 
\begin{align}
&\Delta ={4\pi |M| \over \mathcal{A}}\left(\hat{N}_- +{1\over 2} \right), ~~~\hat{N}_-\equiv \hat{a}_-^{\dag}\hat{a}_-, \notag \\ 
&\hat{a}_-\equiv i\sqrt{\mathcal{A} \over \pi
  |M|}D_{z}^{(b)},~~~\hat{a}_-^{\dag}\equiv i\sqrt{\mathcal{A} \over
  \pi |M|}
D_{\bar{z}}^{(b)},~~~[\hat{a}_-,\hat{a}_-^{\dag}]=1, 
\label{1DHOinQM_-}
\end{align}
with
\begin{align}
\Bigl|0,{j+\alpha_1 \over M},\alpha_{\tau}\Bigr\rangle_{T^2-} \equiv \psi_{T^2-,0}^{(j+\alpha_1,\alpha_{\tau})}(z).
\end{align}
Thus, Kaluza-Klein mode functions are given by
\begin{align}
\psi_{T^2+,n}^{(j+\alpha_1,\alpha_{\tau})}(z)
&\equiv {1\over \sqrt{n!}}(\hat{a}_+^{\dag})^{n}\psi_{T^2+,0}^{(j+\alpha_1,\alpha_{\tau})}(z) \notag \\
&={\mathcal{N} \over \sqrt{n!}}e^{i\pi M
  (z+a_w){\mathrm{Im}(z+a_w)\over \mathrm{Im}\tau}}
\sum_{l=-\infty}^{\infty}e^{i\pi ({j+\alpha_1 \over M}+l)^2M\tau}
e^{2\pi i({j+\alpha_1 \over M}+l)(M(z+a_w)-\alpha_{\tau})} \notag \\
&\hspace{25mm}\times\mathcal{H}_n\left(\sqrt{4\pi
    M\mathrm{Im}\tau}\left({\mathrm{Im}(z+a_w) \over
      \mathrm{Im}\tau}+{j+\alpha_1 \over M}+l\right)\right), 
\notag \\
\psi_{T^2-,n}^{(j+\alpha_1,\alpha_{\tau})}(z)
&={2\over m_n}D_{\bar{z}}^{(b)}\psi_{T^2+,n}^{(j+\alpha_1,\alpha_{\tau})}(z)\hspace{60mm}\mathrm{for}~~M>0, 
\end{align}
\begin{align}
\psi_{T^2-,n}^{(j+\alpha_1,\alpha_{\tau})}(z)
&\equiv {1\over \sqrt{n!}}(\hat{a}_-^{\dag})^{n}\psi_{T^2-,0}^{(j+\alpha_1,\alpha_{\tau})}(z) \notag \\
&={\mathcal{N} \over \sqrt{n!}}e^{i\pi M
  (\bar{z}+\bar{a}_w){\mathrm{Im}(\bar{z}+\bar{a}_w)\over
    \mathrm{Im}\bar{\tau}}} \sum_{l=-\infty}^{\infty}e^{i\pi
  ({j+\alpha_1 \over M}+l)^2M\bar{\tau}}e^{2\pi i({j+\alpha_1 \over
    M}+l)(M(\bar{z}+\bar{a}_w)-\alpha_{\tau})} 
\notag \\
&\hspace{25mm}\times\mathcal{H}_n\left(\sqrt{4\pi
    M\mathrm{Im}\bar{\tau}}\left({\mathrm{Im}(\bar{z}+\bar{a}_w) \over
      \mathrm{Im}\bar{\tau}}+{j+\alpha_1 \over M}+l\right)\right), 
\notag \\
\psi_{T^2+,n}^{(j+\alpha_1,\alpha_{\tau})}(z)
&=-{2\over m_n}D_{z}^{(b)}\psi_{T^2-,n}^{(j+\alpha_1,\alpha_{\tau})}(z)\hspace{56mm}\mathrm{for}~~M<0,
\end{align} 
where $\mathcal{H}_n$ is the Hermite polynomial defined as
\begin{align}
\mathcal{H}_n(x)=(-1)^ne^{x^2\over 2}{d^n \over dx^n}e^{-{x^2\over 2}}.
\end{align}
These Kaluza-Klein mode functions satisfy the orthonormality condition on $T^2$
\begin{align}
\int_{T^2}dzd\bar{z}~\psi_{T^2\pm,m}^{(j+\alpha_1,\alpha_{\tau})}(z)(\psi_{T^2\pm,n}^{(k+\alpha_1,\alpha_{\tau})}(z))^*=\delta_{mn}\delta^{jk}.
\end{align} 

It should be noted that both of the non-zero-mode functions
$\psi_{T^2+,n}^{(j+\alpha_1,\alpha_{\tau})}(z)$ and
$\psi_{T^2-,n}^{(j+\alpha_1,\alpha_{\tau})}(z)$ are well-defined for
$M>0$ and $M<0$, though only
$\psi_{T^2+,0}^{(j+\alpha_1,\alpha_{\tau})}(z)$
$(\psi_{T^2-,0}^{(j+\alpha_1,\alpha_{\tau})}(z))$ 
on the zero-mode functions are well-defined for $M>0~(M<0)$.
The masses squared of $\psi_{T^2\pm,n}^{(j+\alpha_1,\alpha_{\tau})}(z)$ are found to be of the form
\begin{align}
m_n^2={4\pi |M| \over \mathcal{A}}n~~~~~~~\mathrm{for}~~n \in \{0,\mathbb{N}\}.
\label{KK mass}
\end{align}
Note that these are masses squared for spinor fields, 
while eigenvalues of $\Delta$ correspond to 
masses squared for scalar fields as 
$m_n^2 = {4\pi |M| \over \mathcal{A}}(n+\frac{1}{2})$ for $n \in \{0,\mathbb{N}\}$.

\begin{figure}[htbp]
\begin{center}
\hspace{6mm}
\includegraphics[clip,width=75mm]{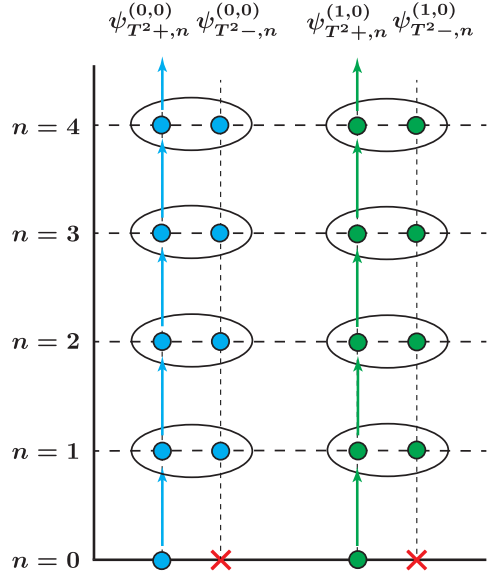}
\end{center}
\caption{The mass spectrum of $\psi_{T^2\pm,n}^{(j,0)}(z)~(j=0,1)$ for $M=2$.
The red crosses mean the absence of zero-mode solutions, and the blue (green) filled circles correspond to a zero mode solution and its Kaluza-Klein modes. 
The blue (green) arrows mean that $\hat{a}^{\dag}$ operates on $n$th
modes $\psi_{T^2+,n}^{(j,0)}(z)$ 
and the next modes $\psi_{T^2+,n+1}^{(j,0)}(z)$ are made by it. 
Two modes in each black oval make a pair to have a mass term.}
\label{fig:ver.torus}
\end{figure}

As an illustrative example, the mass spectra of
$\psi_{T^2\pm,n}^{(j,0)}(z)~(j=0,1)$ for $M=2$ and
$(\alpha_1,\alpha_{\tau})=(0,0)$ are depicted 
in figure~\ref{fig:ver.torus}.
The red crosses mean the absence of zero-mode solutions, and the blue
(green) filled circles correspond to a zero mode and its Kaluza-Klein
modes of 
$\psi_{T^2\pm,n}^{(0,0)}(z)~(\psi_{T^2\pm,n}^{(1,0)}(z))$. 
The blue (green) arrows mean that $\hat{a}^{\dag}$ operates on the $n$th
modes $\psi_{T^2+,n}^{(j,0)}(z)$ and the next modes
$\psi_{T^2+,n+1}^{(j,0)}(z)$ 
are made by it.
Two modes in each black oval, $\psi^{(0,0)}_{T^2+,n}$
and  $\psi^{(0,0)}_{T^2-,n}$ $(\psi^{(1,0)}_{T^2+,n}$
and  $\psi^{(1,0)}_{T^2-,n}$), make a pair to have a
  mass term.

Let us study the masses of Kaluza-Klein modes on $M^4\times T^2/Z_N$.
The masses for $\psi_{T^2/Z_N\pm,n}^{(j+\alpha_1,\alpha_{\tau})}(z)$ are given by 
\begin{align}
\left(
\begin{array}{cc}
-4D_z^{(b)}D_{\bar{z}}^{(b)} & 0 \\
0 & -4D_{\bar{z}}^{(b)}D_{z}^{(b)}
\end{array}
\right)\left(
\begin{array}{c}
\psi_{T^2/Z_N+,n}^{(j+\alpha_1,\alpha_{\tau})}(z) \\ \psi_{T^2/Z_N-,n}^{(j+\alpha_1,\alpha_{\tau})}(z)
\end{array}
\right) =m_n^2\left(
\begin{array}{cc}
\psi_{T^2/Z_N+,n}^{(j+\alpha_1,\alpha_{\tau})}(z) \\ \psi_{T^2/Z_N-,n}^{(j+\alpha_1,\alpha_{\tau})}(z)
\end{array}
\right).
\end{align}
Since $\psi_{T^2/Z_N\pm,0}^{(j+\alpha_1,\alpha_{\tau})}(z)_{\eta}$ are
made by linear combinations of
$\psi_{T^2\pm,0}^{(j+\alpha_1,\alpha_{\tau})}(z)$ 
as we have discussed in section~\ref{sec:Z_Neigenstate}, the Kaluza-Klein
modes on the orbifolds should be made by operating
$(\hat{a}_{\pm}^{\dag})^n$ on 
$\Bigl|0,{j+\alpha_1 \over M},\alpha_{\tau}\Bigr\rangle_{T^2/Z_N\pm,\eta} \equiv \psi_{T^2/Z_N\pm,0}^{(j+\alpha_1,\alpha_{\tau})}(z)_{\eta}$.
Here, we should notice that for $\hat{a}_+^{\dag}$,
$D_{\omega^{k}z}^{(b)}$ are operated on
$\psi_{T^2+,0}^{(j+\alpha_1,\alpha_{\tau})}(\omega^{k} z)$, while for
$\hat{a}_-^{\dag}$, $D_{\bar{\omega}^{k}\bar{z}}^{(b)}$ are operated 
on $\psi_{T^2-,0}^{(j+\alpha_1,\alpha_{\tau})}(\omega^{k} z)$.
We define $\hat{a}_{\omega^{k}\pm}$ and $\hat{a}_{\omega^{k}\pm}^{\dag}$ as
\begin{align}
&\hat{a}_{\omega^{k}+}=\omega^{k}\hat{a}_{+},~~~\hat{a}_{\omega^{k}+}^{\dag}=\bar{\omega}^{k}\hat{a}_{+}^{\dag},\\
&\hat{a}_{\omega^{k}-}=\bar{\omega}^{k}\hat{a}_{-},~~~\hat{a}_{\omega^{k}-}^{\dag}=\omega^{k}\hat{a}_{-}^{\dag},
\end{align}
with $\hat{a}_{1\pm}\equiv \hat{a}_{\pm}$ and $\hat{a}_{1\pm}^{\dag}\equiv \hat{a}_{\pm}^{\dag}$.
Actually, operating $a_{\pm}^{\dag}$ on $\psi_{T^2/Z_N\pm,0}^{(j+\alpha_1,\alpha_{\tau})}(z)_{\eta}$, we obtain the  first Kaluza-Klein modes
\begin{align}
\hat{a}_+^{\dag}\Bigl|0,{j+\alpha_1 \over M},\alpha_{\tau}\Bigr\rangle_{T^2/Z_N+,\eta} 
&=\mathcal{N}^{(j)}_{+,\eta}\sum_{k =0}^{N-1}\bar{\eta}^{k} \hat{a}_+^{\dag}\psi_{T^2+,0}^{(j+\alpha_1,\alpha_{\tau})}(\omega^{k}z) \notag \\
&=\mathcal{N}^{(j)}_{+,\eta}\sum_{k =0}^{N-1}(\omega\bar{\eta})^{k}\hat{a}_{\omega^{k}+}^{\dag}\psi_{T^2+,0}^{(j+\alpha_1,\alpha_{\tau})}(\omega^{k}z) \notag \\
&\equiv \psi_{T^2/Z_N+,1}^{(j+\alpha_1,\alpha_{\tau})}(z)_{\bar{\omega} \eta} ,\\
\hat{a}_-^{\dag}\Bigl|0,{j+\alpha_1 \over M},\alpha_{\tau}\Bigr\rangle_{T^2/Z_N-,\eta} 
&=\mathcal{N}^{(j)}_{-,\eta}\sum_{k =0}^{N-1}\bar{\eta}^{k} \hat{a}_-^{\dag}\psi_{T^2-,0}^{(j+\alpha_1,\alpha_{\tau})}(\omega^{k}z) \notag \\
&=\mathcal{N}^{(j)}_{-,\eta}\sum_{k =0}^{N-1}(\bar{\omega}\bar{\eta})^{k} \hat{a}_{\omega^{k}-}^{\dag}\psi_{T^2-,0}^{(j+\alpha_1,\alpha_{\tau})}(\omega^{k}z) \notag \\
&\equiv \psi_{T^2/Z_N-,1}^{(j+\alpha_1,\alpha_{\tau})}(z)_{\omega \eta}. 
\end{align}  
Thus, the $Z_N$ eigenstate with the eigenvalue $\eta$ at the $n$th
Kaluza-Klein modes is made by operating $a_{+}^{\dag}$ on
$\psi_{T^2/Z_N+,n-1}^{(j+\alpha_1,\alpha_{\tau})}(z)_{\omega \eta}$,
or by operating 
$a_{-}^{\dag}$ on $\psi_{T^2/Z_N-,n-1}^{(j+\alpha_1,\alpha_{\tau})}(z)_{\bar{\omega}\eta}$.
The Kaluza-Klein mode functions are given by
\begin{align}
\psi_{T^2/Z_N+,n}^{(j+\alpha_1,\alpha_{\tau})}(z)_{\bar{\omega}^n\eta}
&\equiv {1\over \sqrt{n!}}(\hat{a}_+^{\dag})^{n}\psi_{T^2/Z_N+,0}^{(j+\alpha_1,\alpha_{\tau})}(z)_{\eta} \notag \\
&=\mathcal{N}^{(j)}_{+,\eta}\sum_{k =0}^{N-1}(\omega^n\bar{\eta})^{k} (\hat{a}_{\omega^{k}+}^{\dag})^n\psi_{T^2-,0}^{(j+\alpha_1,\alpha_{\tau})}(\omega^{k}z) \notag \\
&=\mathcal{N}^{(j)}_{+,\eta}\sum_{k =0}^{N-1}(\omega^n\bar{\eta})^{k} \psi_{T^2-,n}^{(j+\alpha_1,\alpha_{\tau})}(\omega^{k}z), \notag \\
\psi_{T^2/Z_N-,n}^{(j+\alpha_1,\alpha_{\tau})}(z)_{\bar{\omega}^n\eta}
&={2\over m_n}D_{\bar{z}}^{(b)}\psi_{T^2/Z_N+,n}^{(j+\alpha_1,\alpha_{\tau})}(z)_{\bar{\omega}^n\eta} \hspace{20mm}\mathrm{for}~~M>0, 
\end{align}
\begin{align}
\psi_{T^2/Z_N-,n}^{(j+\alpha_1,\alpha_{\tau})}(z)_{\omega^n\eta}
&\equiv {1\over \sqrt{n!}}(\hat{a}_-^{\dag})^{n}\psi_{T^2/Z_N-,0}^{(j+\alpha_1,\alpha_{\tau})}(z)_{\eta} \notag \\
&=\mathcal{N}^{(j)}_{-,\eta}\sum_{k =0}^{N-1}(\bar{\omega}^n\bar{\eta})^{k} (\hat{a}_{\omega^{k}-}^{\dag})^n\psi_{T^2-,0}^{(j+\alpha_1,\alpha_{\tau})}(\omega^{k}z) \notag \\
&=\mathcal{N}^{(j)}_{-,\eta}\sum_{k =0}^{N-1}(\bar{\omega}^n\bar{\eta})^{k} \psi_{T^2-,n}^{(j+\alpha_1,\alpha_{\tau})}(\omega^{k}z) \notag \\
\psi_{T^2/Z_N+,n}^{(j+\alpha_1,\alpha_{\tau})}(z)_{\omega^n\eta}
&=-{2\over m_n}D_{z}^{(b)}\psi_{T^2-,n}^{(j+\alpha_1,\alpha_{\tau})}(z)_{\omega^n\eta} \hspace{20mm}\mathrm{for}~~M<0.
\end{align} 
Then, the Kaluza-Klein modes $\psi_{T^2/Z_N\pm,n}^{(j+\alpha_1,\alpha_{\tau})}(z)_{\eta}$ for ${}^{\forall} \eta$ possess the masses squared
\begin{align}
m_n^2={4\pi |M| \over \mathcal{A}}n~~~~~~~\mathrm{for}~~n \in \{0,\mathbb{N}\}.
\end{align}

\begin{figure}[htbp]
\begin{center}
\includegraphics[width=118mm]{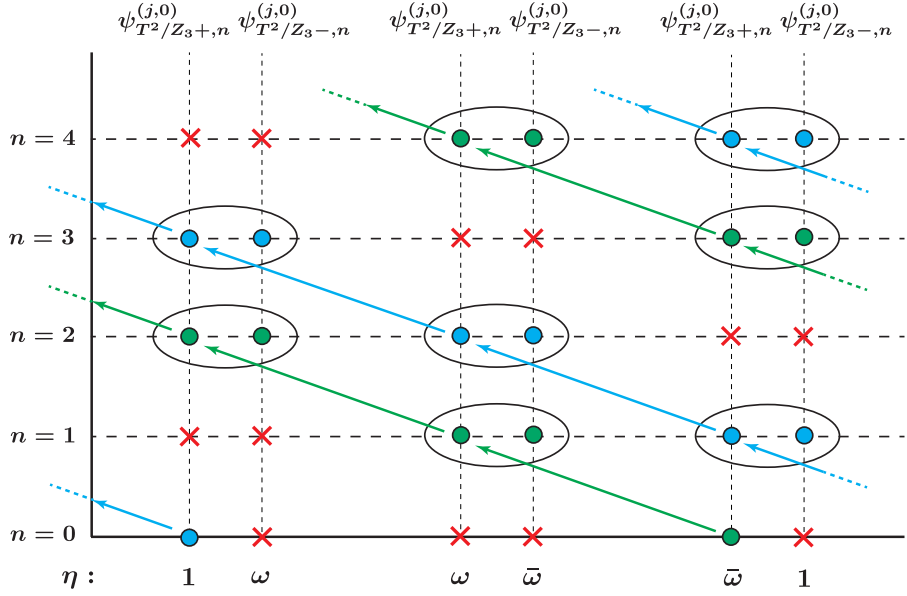}
\end{center}
\caption{The mass spectra of $\psi_{T^2/Z_3\pm,n}^{(j,0)}(z)_{\eta}~(j=0,1)$ for $M=2$ in table~\ref{T2Z3Psi_1e}. 
The red crosses mean the absence of zero-mode solutions and
Kaluza-Klein modes, and the blue (green) filled circles correspond to
a zero mode and
 its Kaluza-Klein modes. 
The blue (green) arrows mean that $\hat{a}^{\dag}$ operates on $n$th
modes $\psi_{T^2/Z_3+,n}^{(j,0)}(z)$ and the next modes
$\psi_{T^2/Z_3+,n+1}^{(j,0)}(z)$
 are made by it. 
Two modes in each black oval make a pair to have a mass term.}
\label{fig:ver.orbifold}
\end{figure}

Here, let us show an illustrative example.
Figure~\ref{fig:ver.orbifold} shows the zero-mode eigenstates
$\psi_{T^2/Z_3+,0}^{(j,0)}(z)_{\eta} ~(j=0,1)$ for $M=2$ in
Table~\ref{T2Z3Psi_1e} and 
its Kaluza-Klein modes.
The meaning of symbols in figure~\ref{fig:ver.orbifold} is the same as in figure~\ref{fig:ver.torus}.
The important difference between figures~\ref{fig:ver.torus} and \ref{fig:ver.orbifold} is  how Kaluza-Klein modes grow up.
In the orbifolds, they grow up as changing the $Z_N$ eigenstates.   

\pagebreak

\section{Conclusions and discussions \label{sec:Conclusion}}

We have studied the $U(1)$ gauge theory on the $T^2/Z_N$ orbifolds with
magnetic fluxes, Scherk-Schwarz phases and Wilson line phases.
We have shown all of the possible 
Scherk-Schwarz and Wilson line phases.
It is remarkable that the allowed Scherk-Schwarz phases 
as well as Wilson line phases depend on the magnitude of 
magnetic flux for the $T^2/Z_3$ and $T^2/Z_6$ orbifolds, 
in particular, whether $M$ is even or odd.
At any rate, the variety of possible Scherk-Schwarz and Wilson line
phases corresponds to the number of fixed points on 
each orbifold with any value of magnetic flux.
Under these backgrounds, we have studied the behavior of 
zero modes.
We have derived the number of zero modes for each 
eigenvalue of the $Z_N$ twist.
This result was obtained by showing explicitly 
and analytically wave functions for some examples 
and also by studying numerically $Z_N$-eigenfunctions for many models.
The exactly same results will be derived by 
another approach for the generic case \cite{Abe:2013}.
The Kaluza-Klein modes were also investigated.

Our results show that one can derive models with 
three generations of matter fermions in various backgrounds, 
i.e., the $T^2/Z_N$ orbifolds for $N=2,3,4,6$ with 
various magnetic fluxes and Scherk-Schwarz phases.
Using these results, one could construct realistic three-generation 
models.
The toroidal compactification can lead to three zero modes 
only when $M=3$, and such a model leads to $\Delta(27)$ flavor 
symmetry \cite{Abe:2009vi,Abe:2009uz,BerasaluceGonzalez:2012vb,
Marchesano:2013ega}.\footnote{A similar flavor symmetry can be 
obtained in heterotic string theory on an orbifold
\cite{Kobayashi:2006wq} (see also \cite{Kobayashi:2004ya,Ko:2007dz}).}
On the other hand, three generations can be realized 
in various orbifold models and that would lead to a rich flavor structure.
Couplings among zero modes in the four-dimensional low energy 
effective field theory are obtained by overlap integrals of 
their wave functions.
Our analysis shows that zero-mode wave functions on the 
orbifold with magnetic flux can be obtained as 
linear combinations of zero-mode wave functions 
on the torus with the same magnetic flux.
Since overlap integrals and couplings of zero-mode 
wave functions on $T^2$ 
were calculated in  \cite{Cremades:2004wa,Abe:2009dr}, 
such couplings can be similarly computed for generic orbifold models.
These analyses on realistic model building for three generations 
and their low energy effective field theory will be 
studied elsewhere.
We have focused on the bulk modes originated from 
higher dimensions.
However, the orbifolds have fixed points.
Then, we can put any localized modes with 
$\delta$-function like profile 
on such fixed points, if that is consistent from the viewpoint 
of four-dimensional field theory.

At any rate, our results can become 
a starting point for these studies.
Also,  our study is applicable to more general twisted orbifold models
in higher-dimensional theory more than six-dimensional one, e.g.,
$T^6/Z_N$, $T^6/(Z_N\times Z_N')$ 
and so on.


\section*{Acknowledgments}

We would like to thank Y. Honma and H. So for discussions.
This work was supported in part by scientific grants from the Ministry of Education, Culture,
Sports, Science and Technology under Grants No.~25$\cdot$3825 (Y.F.), No.~25400252 (T.K.) and No.~20540274 (M.S.).
K.N. is partially supported by funding available from the
Department of Atomic Energy, Government of India for the Regional
Centre for Accelerator-based Particle Physics (RECAPP), Harish-Chandra
Research Institute.

\appendix
\section{Lorentz spinors and gamma matrices\label{LsandGm}}

In our work, the gamma matrices are take as
\begin{align}
&\{\Gamma^M,\Gamma^N\}=2\eta^{MN}=2\mathrm{diag}(+1,-1,-1,-1,-1,-1), \\
&\Gamma^{\mu}=\gamma^{\mu}\otimes \mathbf{1}_{2\times 2}=\left(
\begin{array}{cc}
\gamma^{\mu}& \\
 & \gamma^{\mu}
\end{array}
\right), \notag \\
&\Gamma^{5}=\gamma_{5}\otimes i\sigma_1=\left(
\begin{array}{cc}
&i\gamma_{5} \\
i\gamma_{5} &
\end{array}
\right),
\Gamma^{6}=\gamma_{5}\otimes i\sigma_2=\left(
\begin{array}{cc}
&\gamma_{5} \\
-\gamma_{5} &
\end{array}
\right), \\
&\Gamma_7=\Gamma^0\Gamma^1\Gamma^2\Gamma^3\Gamma^5\Gamma^6=\gamma_5\otimes \sigma_3 =\left(
\begin{array}{cc}
\gamma_5 & \\
 & -\gamma_5
\end{array}
\right) ,~
\{\Gamma_7,\Gamma^M\}=0,
\end{align}
where $M=0,1,2,3,5,6$, $\gamma^{\mu}~(\mu =0,1,2,3)$ is
a four-dimensional gamma matrix,
$\gamma_5=i\gamma^0\gamma^1\gamma^2\gamma^3$ and 
$\sigma_i~(i=1,2,3)$ is a Pauli matrix. 
Furthermore we rewrite the fifth and sixth dimensions as a two-dimensional complex plane such as
\begin{align}
z=x_5+ix_6,~\bar{z}=x_5-ix_6.
\end{align}
Then the gamma matrices and the 
differential operators are given by
\begin{align}
&\Gamma^z \equiv \Gamma^5+i\Gamma^6=\left(
\begin{array}{cc}
 & 2i\gamma_5 \\
0 &  
\end{array}
\right),~
\Gamma^{\bar{z}} \equiv \Gamma^5-i\Gamma^6=\left(
\begin{array}{cc}
 & 0 \\
2i\gamma_5 &  
\end{array}
\right), \\
&\partial_z={1\over 2}(\partial_5 -i\partial_6),~\partial_{\bar{z}}={1\over 2}(\partial_5 +i\partial_6).
\end{align}
Here the translations from ${5,6}$ to ${z,\bar{z}}$ satisfy  
\begin{align}
&A^a=\eta^{ab}A_b,~B_a=\eta_{ab}B^b~(a,b=z,\bar{z})\notag \\
&\hspace{3cm};~
(\eta^{ab})=\left(
\begin{array}{cc}
0&2\\2&0
\end{array}
\right),~
(\eta_{ab})=\left(
\begin{array}{cc}
0&{1\over 2}\\{1\over 2}&0
\end{array}
\right),\notag \\
&A_a={g_a}^{\alpha}A_{\alpha},~B_{\alpha}={(g^{-1})_{\alpha}}^aB_a~(a=z,\bar{z},~\alpha=5,6)\notag \\
&\hspace{3cm};~
g={1\over 2}\left(
\begin{array}{cc}
1&-i\\1&~~i
\end{array}
\right),~
g^{-1}=\left(
\begin{array}{cc}
1&~~1\\i&-i
\end{array}
\right)
=2g^{\dag}.
\end{align}

\section{Redefinition of fields\label{Wlp&alp}}

We consider the relation between the redefinition of  a field
$\Phi(z)$ interacting with a $U(1)$ gauge field $A(z)$ and
Scherk-Schwarz phases
 $\alpha_{1}$ and $\alpha_{\tau}$ which are real numbers.
Here, we denote $\Phi(z)$ and $A(z)$ as $\Phi(z;a_w)$ and $A(z;a_w)$ to emphasize the Wilson line phase $a_w$.
We have already obtained the boundary conditions for $A(z;a_w)$ and $\Phi(z;a_w)$ on the torus with magnetic flux, \begin{align}
&A(z+1;a_w)=A(z;a_w)+d\chi_1 (z+a_w),~~A(z+\tau;a_w)=A(z;a_w) +d\chi_{\tau} (z+a_w), \notag \\
&\Phi(z+1;a_w) =e^{iq\chi_1(z+a_w) +2\pi i \alpha_1}\Phi (z;a_w),~~\Phi(z+\tau;a_w) =e^{iq\chi_{\tau}(z+a_w) +2\pi i \alpha_{\tau}}\Phi(z;a_w),
\end{align}  
with
\begin{align}
&\chi_1(z+a_w)={\pi M\over q\mathrm{Im}\tau}\mathrm{Im}(z+a_w),~~
\chi_{\tau}(z+a_w)={\pi M\over q\mathrm{Im}\tau}\mathrm{Im}[\bar{\tau}(z+a_w)].
\end{align}
Let us redefine $\Phi(z;a_w)$ by
\begin{align}
\Phi(z;a_w)\equiv e^{iq\mathrm{Re}[\bar{\gamma} (z+\tilde{a}_w)]}\tilde{\Phi}(z;\tilde{a}_w),
\label{redefine_Phi}
\end{align}
where $\gamma$ is a complex number, and $\tilde{a}_w$ will be determined below.
With this redefinition, the covariant derivatives for $\Phi$ can be written by
\begin{align}
&(\partial_z -iqA_z(z;a_w))\Phi(z;a_w)=e^{iq\mathrm{Re}[\bar{\gamma}
  (z+\tilde{a}_w)]}(\partial_z
-iq\tilde{A}_z(z;\tilde{a}_w))\tilde{\Phi}(z;\tilde{a}_w), 
\notag \\
&(\partial_{\bar{z}}
-iqA_{\bar{z}}(z;a_w))\Phi(z;a_w)=e^{iq\mathrm{Re}[\bar{\gamma}
  (z+\tilde{a}_w)]}(\partial_{\bar{z}}
-iq\tilde{A}_{\bar{z}}(z;\tilde{a}_w))\tilde{\Phi}
(z;\tilde{a}_w).
\end{align}
Here we defined $\tilde{A}_z$ and $\tilde{A}_{\bar{z}}$ as
$\tilde{A}_z\equiv A_z-\bar{\gamma}/2$ and $\tilde{A}_{\bar{z}}\equiv
A_{\bar{z}}-\gamma/2$, 
respectively.
Then the Wilson line phase for $\tilde{A}$ is given by 
\begin{align}
M\tilde{a}_w\equiv Ma_w+{iq\gamma \mathrm{Im}\tau \over \pi }.
\label{SSphases to Wilson line phase}
\end{align}
We notice that under the transformation $\Phi \to \tilde{\Phi}$ and $A \to \tilde{A}$,
the Lagrangian density $\mathcal{L}$ is invariant, i.e., $\mathcal{L}(A,\Phi)=\mathcal{L}(\tilde{A},\tilde{\Phi})$.
Using 
\begin{align}
\chi_1(z+\tilde{a}_w)= \chi_1(z+a_w) +\mathrm{Re}\gamma ,~~\chi_{\tau}(z+\tilde{a}_w)= \chi_{\tau}(z+a_w) +\mathrm{Re}(\bar{\tau} \gamma),
\end{align}
we can obtain that $\tilde{A}$ and $\tilde{\Phi}$ satisfy  
\begin{align}
&\tilde{A}(z+1;\tilde{a}_w)=\tilde{A}(z;\tilde{a}_w)+d\chi_1(z+\tilde{a}_w),~~\notag \\
&\tilde{A}(z+\tau;\tilde{a}_w)=\tilde{A}(z;\tilde{a}_w)+d\chi_{\tau}(z+\tilde{a}_w),\notag \\
&\tilde{\Phi}(z+1;\tilde{a}_w)=e^{iq\chi_1(z+\tilde{a}_w) +2\pi i\alpha_1 -2iq\mathrm{Re}\gamma }\tilde{\Phi}(z;\tilde{a}_w),~~\notag \\
&\tilde{\Phi}(z+\tau;\tilde{a}_w)=e^{iq\chi_{\tau}(z+\tilde{a}_w) +2\pi i\alpha_{\tau} -2iq\mathrm{Re}(\bar{\tau} \gamma)}\tilde{\Phi}(z;\tilde{a}_w).
\end{align}
If we take $\gamma$ to satisfy $\pi \alpha_1 -q\mathrm{Re}\gamma =0$
and $\pi \alpha_{\tau} -q\mathrm{Re}(\bar{\tau} \gamma) =0$, we obtain
the redefined Wilson line phase $\tilde{a}_w$ and the redefined
Scherk-Schwarz phases $\tilde{\alpha}_1$ and 
$\tilde{\alpha}_{\tau}$,
\begin{align}
M\tilde{a}_w=Ma_w+\alpha_1\tau -\alpha_{\tau},~~\tilde{\alpha}_1=0,~~\tilde{\alpha}_{\tau}=0.
\label{a1=0a2=0condition}
\end{align}
Thus, we can always make the Scherk-Schwarz phases absorbed into the Wilson line phase by the redefinition of fields.
On the other hand, if we take $\gamma$ to satisfy $\tilde{a}_w=0$ in eq.(\ref{SSphases to Wilson line phase}), we obtain
\begin{align}
\tilde{a}_w=0,~~\tilde{\alpha}_1=\alpha_1 +{M\mathrm{Im}a_w \over
  \mathrm{Im}\tau} ,~~\tilde{\alpha}_{\tau}=\alpha_{\tau} 
+{M\mathrm{Im}(\bar{\tau}a_w) \over \mathrm{Im}\tau}.
\label{aw=0condition}
\end{align}
Thus, we can always make the Wilson line phase absorbed into the Scherk-Schwarz phases by the redefinition of fields too.

\section{Example of calculation for linearly independent wave functions\label{Exaple of calculation}}

Let us show the direct analysis with eq.(\ref{psi_T2/ZN+,n}) and its formulae among the $\vartheta$ functions for some examples. 
First, we study the case with $a_w=0$,
$(\alpha_1,\alpha_{\tau})=(0,0)$, $j=0,1$ and $M=2$ on $T^2/Z_4$.
Here, we use $\omega=\tau=i$.
The eigenstates for all $Z_4$ eigenvalues are given by 
\begin{align}
&\psi_{T^2/Z_4+,0}^{(j,0)}(z)_{+1} =\mathcal{N}^{(j)}_{+,+1}\sum_{k =0}^3\psi_{T^2+,0}^{(j,0)}(i^{k}z), ~~\notag \\
&\psi_{T^2/Z_4+,0}^{(j,0)}(z)_{+i} =\mathcal{N}^{(j)}_{+,+i}\sum_{k =0}^3(-i)^{k}\psi_{T^2+,0}^{(j,0)}(i^{k}z), \notag \\
&\psi_{T^2/Z_4+,0}^{(j,0)}(z)_{-1} =\mathcal{N}^{(j)}_{+,-1}\sum_{k =0}^3(-1)^{k}\psi_{T^2+,0}^{(j,0)}(i^{k}z), ~~\notag \\
&\psi_{T^2/Z_4+,0}^{(j,0)}(z)_{-i} =\mathcal{N}^{(j)}_{+,-i}\sum_{k =0}^3i^{k}\psi_{T^2+,0}^{(j,0)}(i^{k}z).
\end{align}
For convenience, we rewrite $\psi_{T^2+,0}^{(j,0)}(z)$ as
\begin{align}
&\psi_{T^2+,0}^{(0,0)}(z)
=\mathcal{N}  e^{2\pi iz\mathrm{Im}z} \vartheta_3 (2z,2i), ~~~
\psi_{T^2+,0}^{(1,0)}(z)
=\mathcal{N}  e^{2\pi iz\mathrm{Im}z} \vartheta_2 (2z,2i),
\end{align}
with the elliptic theta functions, which are defined as 
\begin{align}
&\vartheta_1(\nu,\tau) = \vartheta \left[
\begin{array}{c}
{1\over 2}\\ {1\over 2}
\end{array}
\right] (\nu,\tau), ~~~
\vartheta_2(\nu,\tau) =\vartheta \left[
\begin{array}{c}
{1\over 2}\\ 0
\end{array}
\right] (\nu,\tau), \notag \\
&\vartheta_3(\nu,\tau) =\vartheta \left[
\begin{array}{c}
0\\ 0
\end{array}
\right] (\nu,\tau), ~~~
\vartheta_4(\nu,\tau) =\vartheta \left[
\begin{array}{c}
0\\ {1\over 2}
\end{array}
\right] (\nu,\tau).
\end{align} 
The following formulae:
\begin{align}
&\vartheta_1(\nu,\tau) \vartheta_1(\mu,\tau) =
\vartheta_3(\nu+\mu,2\tau) \vartheta_2(\nu-\mu,2\tau) -
\vartheta_2(\nu+\mu,2\tau) \vartheta_3(\nu-\mu,2\tau), 
\notag \\
&\vartheta_3(\nu,\tau) \vartheta_3(\mu,\tau) =
\vartheta_3(\nu+\mu,2\tau) \vartheta_3(\nu-\mu,2\tau) +
\vartheta_2(\nu+\mu,2\tau) \vartheta_2(\nu-\mu,2\tau), 
\notag \\
&\vartheta_2(2\nu,2\tau) = {1\over \vartheta_4(0,2\tau)} \vartheta_1({1\over 4} -\nu,\tau)\vartheta_1({1\over 4} +\nu,\tau), \notag \\
&\vartheta_3(2\nu,2\tau) = {1\over \vartheta_4(0,2\tau)} \vartheta_3({1\over 4} -\nu,\tau)\vartheta_3({1\over 4} +\nu,\tau),
\label{eq_theta_relation}
\end{align} 
are useful.
To understand the number of zero-mode eigenstates, we need to rewrite
$\psi_{T^2+,0}^{(0,0)}(iz)$ and $\psi_{T^2+,0}^{(1,0)}(iz)$ by 
$\psi_{T^2+,0}^{(0,0)}(z)$ and $\psi_{T^2+,0}^{(1,0)}(z)$.
Namely, we need to calculate $\vartheta_2(2iz,2i)$ and $\vartheta_3(2iz,2i)$.
Using the relations (\ref{eq_theta_relation}), we can obtain
\begin{align}
\vartheta_2(2iz,2i)
&={1\over \vartheta_4(0,2i)} \vartheta_1({1\over 4} -iz,i)\vartheta_1({1\over 4} +iz,i) \notag \\
&=-{1\over \vartheta_4(0,2i)} e^{-{\pi \over 8}}e^{2\pi z^2}\vartheta_1({i\over 4} +z,i)\vartheta_1({i\over 4} -z,i) \notag \\
&=-{1\over \vartheta_4(0,2i)} e^{-{\pi \over 8}}e^{2\pi
  z^2}\left[\vartheta_3({i\over 2}
  ,2i)\vartheta_2(2z,2i)-\vartheta_2({i\over 2} ,2i)
\vartheta_3(2z,2i)\right] \notag \\
\vartheta_3(2iz,2i)
&={1\over \vartheta_4(0,2i)} \vartheta_3({1\over 4} -iz,i)\vartheta_3({1\over 4} +iz,i)\notag \\
&={1\over \vartheta_4(0,2i)} e^{-{\pi \over 8}}e^{2\pi z^2}\vartheta_3({i\over 4} +z,i)\vartheta_3({i\over 4} -z,i) \notag \\
&={1\over \vartheta_4(0,2i)} e^{-{\pi \over 8}}e^{2\pi
  z^2}\left[\vartheta_3({i\over 2}
  ,2i)\vartheta_3(2z,2i)+\vartheta_2({i\over 2} ,2i)
\vartheta_2(2z,2i)\right].
\end{align} 
{}From these results, we can rewrite $\psi_{T^2+,0}^{(j,0)}(iz)$ by
\begin{align}
\psi_{T^2+,0}^{(0,0)}(iz)
&=\mathcal{N}e^{2\pi i(iz)\mathrm{Im}(iz)} \vartheta_3 (2iz,2i), \notag \\
&=\mathcal{N}e^{2\pi i(iz)\mathrm{Im}(iz)}{1\over \vartheta_4(0,2i)}
e^{-{\pi \over 8}}e^{2\pi z^2}\left[\vartheta_3({i\over 2}
  ,2i)\vartheta_3(2z,2i)+\vartheta_2({i\over 2}
  ,2i)\vartheta_2(2z,2i)\right] 
\notag \\
&=\mathcal{N}e^{-{\pi \over 8}}e^{2\pi
  iz\mathrm{Im}z}\left[{\vartheta_3({i\over 2} ,2i) \over
    \vartheta_4(0,2i)} \vartheta_3(2z,2i) +{\vartheta_2({i\over 2}
    ,2i) \over \vartheta_4(0,2i)} \vartheta_2(2z,2i)\right] 
\notag \\
&=e^{-{\pi \over 8}}\left[{\vartheta_3({i\over 2} ,2i) \over
    \vartheta_4(0,2i)} \psi_{T^2+,0}^{(0,0)}(z) +{\vartheta_2({i\over
      2} ,2i) \over \vartheta_4(0,2i)} \psi_{T^2+,0}^{({1\over
      2},0)}(z)\right] 
\notag \\
&={1\over \sqrt{2}}\left(\psi_{T^2+,0}^{(0,0)}(z)+\psi_{T^2+,0}^{(1,0)}(z)\right), \notag \\
\psi_{T^2+,0}^{(1,0)}(iz)
&=\mathcal{N}e^{2\pi i(iz)\mathrm{Im}(iz)} \vartheta_2 (2iz,2i), \notag \\
&=\mathcal{N}e^{2\pi i(iz)\mathrm{Im}(iz)}{-1\over \vartheta_4(0,2i)}
e^{-{\pi \over 8}}e^{2\pi z^2}\left[\vartheta_3({i\over 2}
  ,2i)\vartheta_2(2z,2i)-\vartheta_2({i\over 2}
  ,2i)\vartheta_3(2z,2i)\right] 
\notag \\
&=-\mathcal{N}e^{-{\pi \over 8}}e^{2\pi
  iz\mathrm{Im}z}\left[{\vartheta_3({i\over 2} ,2i) \over
    \vartheta_4(0,2i)} \vartheta_2(2z,2i) -{\vartheta_2({i\over 2}
    ,2i) \over \vartheta_4(0,2i)} \vartheta_3(2z,2i)\right] 
\notag \\
&= -e^{-{\pi \over 8}}\left[{\vartheta_3({i\over 2} ,2i) \over
    \vartheta_4(0,2i)} \psi_{T^2+,0}^{({1\over 2},0)}(z)
  -{\vartheta_2({i\over 2} ,2i) \over \vartheta_4(0,2i)}
  \psi_{T^2+,0}^{(0,0)}(z)\right] 
\notag \\
&={1\over \sqrt{2}}\left(\psi_{T^2+,0}^{(0,0)}(z)-\psi_{T^2+,0}^{(1,0)}(z)\right).
\end{align} 
Finally, we can obtain the zero-mode eigenstates for all $Z_4$ eigenvalues
\begin{align}
&\psi_{T^2/Z_4+,0}^{(0,0)}(z)_{+1} 
=\mathcal{N}^{(0)}_{+,+1} \left(\psi_{T^2+,0}^{(0,0)}(z)+(\sqrt{2}-1)\psi_{T^2+,0}^{(1,0)}(z)\right), \notag \\
&\psi_{T^2/Z_4+,0}^{(1,0)}(z)_{+1} 
=\mathcal{N}^{(1)}_{+,+1} \left(\psi_{T^2+,0}^{(0,0)}(z)+(\sqrt{2}-1)\psi_{T^2+,0}^{(1,0)}(z)\right), \notag \\
&\psi_{T^2/Z_4+,0}^{(0,0)}(z)_{-1} 
=\mathcal{N}^{(0)}_{+,-1} \left(\psi_{T^2+,0}^{(0,0)}(z)-(\sqrt{2}+1)\psi_{T^2+,0}^{(1,0)}(z)\right), \notag \\
&\psi_{T^2/Z_4+,0}^{(1,0)}(z)_{-1} 
=\mathcal{N}^{(1)}_{+,-1} \left(\psi_{T^2+,0}^{(0,0)}(z)-(\sqrt{2}+1)\psi_{T^2+,0}^{(1,0)}(z)\right), \notag \\
&\psi_{T^2/Z_4+,0}^{(0,0)}(z)_{\pm i}=\psi_{T^2/Z_4+,0}^{(1,0)}(x,z)_{\pm i}=0.
\end{align}
The number of independent wave functions is equal to two.
Thus, the total number of zero-mode eigenstates is equal to two, and corresponds to the magnitude of magnetic flux $|M|$.
This result corresponds to the case for $M=2$ in Table \ref{T2Z4Psi_1}.



\begin{thebibliography}{99}


\bibitem{Bachas:1995ik}
  C.~Bachas,
  arXiv:hep-th/9503030;
%
  R.~Blumenhagen, L.~Goerlich, B.~Kors and D.~Lust,
  JHEP {\bf 0010}, 006 (2000)
  [arXiv:hep-th/0007024];
%
  C.~Angelantonj, I.~Antoniadis, E.~Dudas and A.~Sagnotti,
  Phys. Lett. {\bf B489}, 223 (2000)
  [arXiv:hep-th/0007090];
%
  R.~Blumenhagen, B.~Kors and D.~Lust,
  JHEP {\bf 0102}, 030 (2001)
  [arXiv:hep-th/0012156].


\bibitem{Cremades:2004wa}
  D.~Cremades, L.~E.~Ibanez and F.~Marchesano,
  JHEP {\bf 0405}, 079 (2004)
  [hep-th/0404229].




\bibitem{Blumenhagen:2005mu}
  R.~Blumenhagen, M.~Cvetic, P.~Langacker and G.~Shiu,
Ann. Rev. Nucl. Part. Sci. {\bf 55}, 71 (2005)
  [arXiv:hep-th/0502005];
%
  R.~Blumenhagen, B.~Kors, D.~Lust and S.~Stieberger,
  Phys. Rept. {\bf 445}, 1 (2007)
  [arXiv:hep-th/0610327].

\bibitem{Cvetic:2003ch} 
  M.~Cvetic and I.~Papadimitriou,
  Phys.\ Rev.\ D {\bf 68}, 046001 (2003)
  [Erratum-ibid.\ D {\bf 70}, 029903 (2004)]
  [hep-th/0303083];
  S.~A.~Abel and A.~W.~Owen,
  Nucl.\ Phys.\ B {\bf 663}, 197 (2003)
  [hep-th/0303124];
%
  D.~Cremades, L.~E.~Ibanez and F.~Marchesano,
  JHEP {\bf 0307}, 038 (2003)
  [hep-th/0302105];
  G.~Honecker and J.~Vanhoof,
  JHEP {\bf 1204}, 085 (2012)
  [arXiv:1201.3604 [hep-th]].


\bibitem{Fujimoto:2012} 
  Y.~Fujimoto, T.~Nagasawa, K.~Nishiwaki and M.~Sakamoto,
  Prog. Theor. Exp. Phys. {\bf 023B07} (2013) 
  [arXiv:1209.5150 [hep-ph]];
%
  Y.~Fujimoto, K.~Nishiwaki and M.~Sakamoto,
  arXiv:1301.7253 [hep-ph].


\bibitem{Abe:2012fj} 
  H.~Abe, T.~Kobayashi, H.~Ohki, A.~Oikawa and K.~Sumita,
  Nucl.\ Phys.\ B {\bf 870}, 30 (2013)
  [arXiv:1211.4317 [hep-ph]].


\bibitem{Abe:2009dr} 
  H.~Abe, K.~-S.~Choi, T.~Kobayashi and H.~Ohki,
  JHEP {\bf 0906}, 080 (2009) 
  [arXiv:0903.3800 [hep-th]].  


\bibitem{Abe:2009vi}
  H.~Abe, K.~-S.~Choi, T.~Kobayashi and H.~Ohki,
  Nucl. Phys. {\bf B820}, 317 (2009)
  [arXiv:0904.2631 [hep-ph]].


\bibitem{Abe:2009uz}
   H.~Abe, K.~-S.~Choi, T.~Kobayashi and H.~Ohki,
  Phys. Rev. {\bf D80}, 126006 (2009)
  [arXiv:0907.5274 [hep-th]]; 
%
  Phys. Rev. {\bf D81}, 126003 (2010)
  [arXiv:1001.1788 [hep-th]].


\bibitem{BerasaluceGonzalez:2012vb} 
  M.~Berasaluce-Gonzalez, P.~G.~Camara, F.~Marchesano, D.~Regalado and A.~M.~Uranga,
  JHEP {\bf 1209}, 059 (2012)
  [arXiv:1206.2383 [hep-th]].


\bibitem{Marchesano:2013ega} 
  F.~Marchesano, D.~Regalado and L.~Vazquez-Mercado,
  arXiv:1306.1284 [hep-th].

\bibitem{Honecker:2013hda} 
  G.~Honecker and W.~Staessens,
  JHEP {\bf 1310}, 146 (2013)
  [arXiv:1303.4415 [hep-th]].


\bibitem{Hamada:2012wj} 
  Y.~Hamada and T.~Kobayashi,
  Prog. Theor. Phys. {\bf 128}, 903 (2012)
  [arXiv:1207.6867 [hep-th]]. 


\bibitem{Sakamoto:2003} 
  M.~Sakamoto and S.~Tanimura,
  J. Math. Phys. {\bf 44}, 5042 (2003)
  [hep-th/0306006].  


\bibitem{Antoniadis:2004pp} 
  I.~Antoniadis and T.~Maillard,
  Nucl. Phys. {\bf B716}, 3 (2005)
  [hep-th/0412008];
%
  I.~Antoniadis, A.~Kumar and B.~Panda,
  Nucl. Phys. {\bf B823}, 116 (2009)
  [arXiv:0904.0910 [hep-th]].  


\bibitem{Choi:2009pv} 
  K.~-S.~Choi, T.~Kobayashi, R.~Maruyama, M.~Murata, Y.~Nakai, H.~Ohki and M.~Sakai,
  Eur. Phys. J. {\bf C67}, 273 (2010)
  [arXiv:0908.0395 [hep-ph]].  
%
%
  T.~Kobayashi, R.~Maruyama, M.~Murata, H.~Ohki and M.~Sakai,
  JHEP {\bf 1005}, 050 (2010)
  [arXiv:1002.2828 [hep-ph]].  

\bibitem{DiVecchia:2011mf}
  P.~Di Vecchia, R.~Marotta, I.~Pesando and F.~Pezzella,
  J. Phys. {\bf A44}, 245401 (2011)
  [arXiv:1101.0120 [hep-th]].


\bibitem{Abe:2012ya}
  H.~Abe, T.~Kobayashi, H.~Ohki and K.~Sumita,
  Nucl. Phys. {\bf B863}, 1 (2012)
  [arXiv:1204.5327 [hep-th]].


\bibitem{DeAngelis:2012jc} 
  L.~De Angelis, R.~Marotta, F.~Pezzella and R.~Troise,
  JHEP {\bf 1210}, 052 (2012)
  [arXiv:1206.3401 [hep-th]].


\bibitem{Abe:2013bba} 
  H.~Abe, T.~Kobayashi, H.~Ohki, K.~Sumita and Y.~Tatsuta,
  arXiv:1307.1831 [hep-th].


\bibitem{Abe:2008fi}
  H.~Abe, T.~Kobayashi and H.~Ohki,
  JHEP {\bf 0809}, 043 (2008)
  [arXiv:0806.4748 [hep-th]].


\bibitem{Abe:2008sx}
  H.~Abe, K.~-S.~Choi, T.~Kobayashi and H.~Ohki,
  Nucl. Phys. {\bf B814}, 265 (2009)
  [arXiv:0812.3534 [hep-th]].


\bibitem{Nibbelink:2012de} 
  S.~Groot Nibbelink and P.~K.~S.~Vaudrevange,
  arXiv:1212.4033 [hep-th].  


\bibitem{Fujimoto:2013xha} 
  Y.~Fujimoto, T.~Kobayashi, T.~Miura, K.~Nishiwaki and M.~Sakamoto,
  Phys. Rev. {\bf D87}, 086001 (2013)
  [arXiv:1302.5768 [hep-th]].


\bibitem{Dixon:1985jw} 
  L.~J.~Dixon, J.~A.~Harvey, C.~Vafa and E.~Witten,
  Nucl.\ Phys.\ B {\bf 261}, 678 (1985);
  Nucl.\ Phys.\ B {\bf 274}, 285 (1986).


\bibitem{Katsuki:1989bf} 
  Y.~Katsuki, Y.~Kawamura, T.~Kobayashi, N.~Ohtsubo, Y.~Ono and K.~Tanioka,
  Nucl.\ Phys.\ B {\bf 341}, 611 (1990).


\bibitem{Kobayashi:1991rp} 
  T.~Kobayashi and N.~Ohtsubo,
  Int.\ J.\ Mod.\ Phys.\ A {\bf 9}, 87 (1994).


\bibitem{Choi:2006qh} For example, see for a review 
  K.~-S.~Choi and J.~E.~Kim,
Lect. Notes Phys. {\bf 696}, 1 (2006). 


\bibitem{Kawamura:2008mz} 
  Y.~Kawamura, T.~Kinami and T.~Miura,
  Prog.\ Theor.\ Phys.\  {\bf 120}, 815 (2008)
  [arXiv:0808.2333 [hep-ph]].


\bibitem{Kawamura:2009sa} 
  Y.~Kawamura and T.~Miura,
  Prog.\ Theor.\ Phys.\  {\bf 122}, 847 (2010)
  [arXiv:0905.4123 [hep-th]].


\bibitem{Kawamura:2007} 
  Y.~Kawamura, T.~Kinami and K.~Y.~Oda,
  Phys. Rev.  {\bf D76}, 035001 (2007)
  [hep-ph/0703195].


\bibitem{Kawamura:2009} 
  Y.~Kawamura and T.~Miura,
  Phys. Rev. {\bf D81}, 075011 (2010)
  [arXiv:0912.0776 [hep-ph]].  


\bibitem{Goto:2013jma} 
  Y.~Goto, Y.~Kawamura and T.~Miura,
  Phys. Rev. {\bf D88}, 055016 (2013)
  [arXiv:1307.2631 [hep-ph]].


\bibitem{Scherk:1978ta} 
  J.~Scherk and J.~H.~Schwarz,
  Phys.\ Lett.\ B {\bf 82}, 60 (1979);
%
  Nucl.\ Phys.\ B {\bf 153}, 61 (1979).


\bibitem{Ibanez:1986tp} 
  L.~E.~Ibanez, H.~P.~Nilles and F.~Quevedo,
  Phys.\ Lett.\ B {\bf 187}, 25 (1987).


\bibitem{Kobayashi:1990mi} 
  T.~Kobayashi and N.~Ohtsubo,
  Phys.\ Lett.\ B {\bf 257}, 56 (1991).


\bibitem{Angelantonj:2005hs} 
  C.~Angelantonj, M.~Cardella and N.~Irges,
  Nucl.\ Phys.\ B {\bf 725}, 115 (2005)
  [hep-th/0503179].


\bibitem{Blumenhagen:2005tn} 
  R.~Blumenhagen, M.~Cvetic, F.~Marchesano and G.~Shiu,
  JHEP {\bf 0503}, 050 (2005)
  [hep-th/0502095].

\bibitem{Angelantonj:2009yj} 
  C.~Angelantonj, C.~Condeescu, E.~Dudas and M.~Lennek,
  Nucl.\ Phys.\ B {\bf 818}, 52 (2009)
  [arXiv:0902.1694 [hep-th]];
%
  S.~Forste and G.~Honecker,
  JHEP {\bf 1101}, 091 (2011)
  [arXiv:1010.6070 [hep-th]].

\bibitem{Hashimoto:1997gm} 
  A.~Hashimoto and W.~Taylor,
  Nucl.\ Phys.\ B {\bf 503}, 193 (1997)
  [hep-th/9703217].


\bibitem{Abe:2013} 
T.~-h.~Abe,  Y.~Fujimoto, T.~Kobayashi, T.~Miura, K.~Nishiwaki and M.~Sakamoto,
in preparation.


\bibitem{Kobayashi:2006wq} 
  T.~Kobayashi, H.~P.~Nilles, F.~Ploger, S.~Raby and M.~Ratz,
  {\it Nucl. Phys.} {\bf B768} (2007) 135
  [hep-ph/0611020].


\bibitem{Kobayashi:2004ya} 
  T.~Kobayashi, S.~Raby and R.~-J.~Zhang,
  Nucl.\ Phys.\ B {\bf 704}, 3 (2005)  [hep-ph/0409098]. 


\bibitem{Ko:2007dz} 
  P.~Ko, T.~Kobayashi, J.~-h.~Park and S.~Raby,
  Phys.\ Rev.\ D {\bf 76}, 035005 (2007)  
  [Erratum-ibid.\ D {\bf 76}, 059901 (2007)] 
  [arXiv:0704.2807[hep-ph]].  



\end{thebibliography}
\end{document}